\documentclass[aps, prd, twocolumn, showpacs, superscriptaddress, groupedaddress]{revtex4} 
\usepackage{graphicx}    
\usepackage{amssymb}
\usepackage{amsmath}
\usepackage{subfigure}
\usepackage[pagebackref=false, colorlinks=true]{hyperref}
\hypersetup{
linkcolor=blue,     
citecolor=blue,     
urlcolor=blue}   
\usepackage{soul}
\usepackage{amsmath}
\begin{document}
\title{The equilibrium condition in gravitational collapse and its application to a cosmological scenario}
\author{Dipanjan Dey}
\email{deydipanjan7@gmail.com}
\affiliation{Department of Mathematics and Statistics,
Dalhousie University,
Halifax, Nova Scotia,
Canada B3H 3J5}
\author{N. T. Layden}
\email{nicholas.layden@dal.ca }
\affiliation{Department of Mathematics and Statistics,
Dalhousie University,
Halifax, Nova Scotia,
Canada B3H 3J5}
\author{A. A. Coley}
\email{Alan.Coley@dal.ca }
\affiliation{Department of Mathematics and Statistics,
Dalhousie University,
Halifax, Nova Scotia,
Canada B3H 3J5}
\author{Pankaj S. Joshi}
\email{psjcosmos@gmail.com}
\affiliation{International Centre for Space and Cosmology, School of Arts and Sciences, Ahmedabad University, Ahmedabad, GUJ 380009, India}
\date{\today}

\begin{abstract}
We discuss the equilibrium conditions of the gravitational collapse of a spherically symmetric matter cloud. We analyze the spinor structure of a general collapsing space-time and redefine the equilibrium conditions by using Cartan scalars. We qualitatively investigate the equilibrium configuration of a two-fluid system consisting of a dust-like fluid and a fluid with a negative equation of state. We use our results to investigate certain cosmological scenarios where dark energy can cluster inside the over-dense regions of dark matter and together reaches a stable configuration. We compare the outcomes of our work with existing work where the virialization technique is used to stabilize the two-fluid system.\\
$\textbf{key words}$: Gravitational collapse, Newman-Penrose formalism, Cartan scalars, Structure formation, Dark matter, Dark energy.

\end{abstract}
\maketitle

\section{Introduction}
In cosmological structure formation, a frequently used technique to explain the stability of mega-structures is virialization. A gravitationally collapsing cloud of particles reaches a stable configuration when it virializes. The virial theorem in the Newtonian framework relates the average kinetic energy and the total potential energy of a stable system consisting of $N$ number of particles bounded by some potential energy. A system reaches a virialization state when the particles of the system have some dynamics or kinetic energy but overall the system is static. In the virial theorem a scalar quantity $\mathcal{G}$ is defined as
\begin{equation}
\mathcal{G}=\sum_{i=1}^{N} {\bf p}_i\cdot{\bf r}_i\, ,
\end{equation}
where a system of $N$ particles is considered and ${\bf p}_i$ and ${\bf r}_i$ are the momentum and position vector of $i$th particle, respectively. The quantity $\mathcal{G}$ is known as ``Virial". This scalar quantity can be derived from the moment of inertia of a system of particles. The moment of inertia of a system of $N$ particles can be written as
\begin{eqnarray}
I=\sum_{i=1}^{N} m_i r_i^2\,\, ,
\end{eqnarray}
where $m_i$ is the mass of the $i$th particle.
Therefore, we can write 
\begin{equation}
    \frac12 \frac{dI}{dt}=\mathcal{G}.
\end{equation} 
Now, the derivative of $\mathcal{G}$ with respect to time can be written as
\begin{equation}
\frac12 \frac{d^2I}{dt^2}=\frac{d\mathcal{G}}{dt}=2T+\sum_{i=1}^{N}{\bf F}_i\cdot{\bf r}_i\, ,
\label{dGdt}
\end{equation} 
where ${\bf F}_i$ is net force on the $i$th particle, and $T$ is the total kinetic energy of the system. As ${\bf F}_i$ is the total force on the $i$th particle, it can be written as the sum of all forces on $i$th particle from the other $j$ particles in the system
\begin{equation}
{\bf F}_i=\sum_j^N{\bf F}_{ji}\, .
\end{equation}
In terms of a spherically symmetric potential, the above equation can be written as
\begin{equation}
{\bf F}_i=-\sum_j^N\frac{dV}{dr}\left(\frac{{\bf r}_j-{\bf r}_i}{r_{ij}}\right)\, ,
\end{equation}
where $V$ and $r_{ij}$ are the potential energy and distance between the $i$th particle and $j$th particle, respectively. One can introduce a time-average of $\frac{d\mathcal{G}}{dt}$ or $\frac{d^2I}{dt^2}$ in the following way
\begin{equation}
\frac12\Bigg\langle\frac{d^2I}{dt^2}\Bigg\rangle=\Bigg\langle \frac{d\mathcal{G}}{dt}\Bigg \rangle=\frac{1}{\tau}\int_{0}^{\tau} \frac{d\mathcal{G}}{dt}dt\, ,
\end{equation}
where the time averaging is done over a time $\tau$. Now, for a bound system, every parameter of the system has a finite value and the virial theorem states that the time average value of $\frac{d\mathcal{G}}{dt}$ or $\frac{d^2I}{dt^2}$  over a large time $\tau$ is almost zero: 
\begin{equation}
\frac12\Bigg\langle\frac{d^2I}{dt^2}\Bigg\rangle_{\tau\rightarrow\infty}=\Bigg\langle \frac{d\mathcal{G}}{dt}\Bigg \rangle_{\tau\rightarrow\infty}\approx 0\,\, .
\label{inertia1}
\end{equation}
So, after a large time, the bound system achieves a virial equilibrium state when the following condition holds (from Eq.~(\ref{dGdt})):
\begin{equation}
\langle T\rangle_{\tau\rightarrow\infty} = -\frac12\Bigg\langle \sum_{i=1}^{N}{\bf F}_i\cdot{\bf r}_i  \Bigg\rangle_{\tau\rightarrow\infty}\, .
\end{equation}
 There are primarily four processes by which a collapsing system consisting only of gravitationally interacting particles can achieve a virialized state: violent relaxation, phase mixing, chaotic mixing, and Landau damping \cite{Lynden-Bell67, Merritt99}. In the standard model of structure formation, the primordial homogeneous over-dense regions of dark matter formed due to the quantum fluctuations in the inflationary era had an initial expansion phase and after a certain turnaround radius, the over-dense regions started collapsing due to its own gravitational pull and, ultimately, virialized to form the stable structures. This model of structure formation is known as the top-hat model. In the standard spherical top-hat collapse
model \cite{GunnGott72}, the metric of the over-dense sub-universe is given by the closed Friedmann-Lemaitre-Robertson-Walker (FLRW) model
\begin{eqnarray}
 ds^2 = - dt^2 + {a^2(t)\over 1- r^2 }dr^2 + a^2(t) r^2 d\Omega^2\,,
\label{FLRWMetric}
\end{eqnarray}
where $a(t)$ is the scale factor for the over-dense region specified by
the closed FLRW space-time. The positive curvature constant set to unity indicates the over-density of that particular region with respect to the background universe. The valid intervals of co-ordinates of closed FLRW space-time are  $0\le r \le 1$, $0 \le \theta \le
\pi$ and $0\le \phi < 2\pi$. In the top-hat collapse model, the over-dense region expands first and at some time it starts collapsing. The simplest metric for describing this type of dynamics is the above-mentioned closed FLRW metric. According to the $\Lambda$CDM model of dark matter, the equation of state of the dark matter is zero, i.e it is a dust-like fluid. Therefore, in the top-hat model, the collapsing fluid is considered as dust. Now, it is well known that the collapse of homogeneous dust (namely Oppenheimer-Snyder-Datt collapse) always forms a black hole. However, to model a stable structure, in the top-hat collapse model, the above-mentioned Newtonian virialization technique is used in an ad-hoc way. There are various attempts to describe the general relativistic interpretation of virialization \cite{virialGR1, virialGR2, virialGR3, virialGR4}; however, a lot of development is needed in this direction. 

In this paper, in the realm of general relativity, we discuss an equilibrium configuration that can be achieved by a collapsing object under certain conditions. The equilibrium conditions discussed here were first proposed in \cite{JMN11, Joshi:2013dva}. Here, we write down those conditions in a covariant and frame-independent way. We analyze the spinor structure of the spherically symmetric collapsing spacetime and define the equilibrium conditions using the Cartan scalars. The motivation behind defining the equilibrium conditions by Cartan scalars is to express the equilibrium state of a collapsing system in a coordinate independent way. It should be noted that in this paper we do not claim that the equilibrium configuration is the virialized state of the collapsing object rather, here, we are interested in showing the difference between the end stable states achieved by satisfying the equilibrium conditions and virialized conditions. In order to do that we introduce pressure inside the collapsing matter since, as discussed before, there exists no equilibrium state of a collapsing dust-like matter. We show that with non-zero internal pressure, the equilibrium configuration can be achieved in a large co-moving time. We use the general relativistic equilibrium conditions to  model certain cosmological scenarios. Since dark matter is generally considered pressure-less fluid, we introduce pressure by considering clustered dark energy. There are a large number of articles \cite{Lahav:1991wc, Steinh, Shapiro, Basilakos:2003bi, Caldwell:2003vq, Horellou:2005qc, Maor:2005hq, Percival:2005vm, mota2006, Wang:2005ad, Basilakos:2006us, Maor:2006rh, Basilakos:2009mz, Basilakos:2010rs, Leewang, Chang:2017vhs} in which the effect of clustered dark energy on the virialization of dark matter in the Hubble scale and the galaxy cluster scale is investigated, where the authors use the Newtonian virialization technique to stabilize the collapsing system consisting of pressure-less dark matter and clustered dark energy. The introduction of clustered dark energy in the over-dense region of dark matter modifies the standard model of the spherical top-hat collapse. In the present paper, we use general relativistic equilibrium conditions to investigate the stable end state of the above-mentioned two-fluid system. Here, we qualitatively analyze the equilibrium configuration of the two-fluid system. We do not compare our results with any existing observational data and leave it for future investigation.

The paper is organized as follows: we begin Sec.~(\ref{nondust}) with a brief review of equilibrium conditions in general relativity and then, in that section, we construct a covariant and frame independent counterpart of the equilibrium conditions. In Sec.~(\ref{tophat}), we briefly discuss the standard top-hat collapse model and review the work on its modifications by introducing clustered dark energy in the over-dense region of dark matter. In Sec.~(\ref{themodel}), we discuss the equilibrium configuration of the two-fluid consisting of dark matter and clustered dark energy. Finally, in Sec.~(\ref{conclude}), we conclude with a discussion and summary of our results. Throughout the paper, we use a system of units in which the velocity of light and the universal gravitational constant (multiplied by $8\pi$), are both set equal to unity.
\section{Equilibrium conditions in gravitational collapse}
\label{nondust}
In this section, we first briefly review the equilibrium conditions in gravitational collapse \cite{JMN11, Joshi:2013dva}, and then show that if the Eq.~(\ref{inertia1}) is written in a general relativistic covariant way, at least for the homogeneous configuration, one can show that the modified covariant form of that equation implies the equilibrium condition in gravitational collapse. In the last subsection, we discuss the spinor structure of the general collapsing space-time and redefine the equilibrium condition by using Cartan scalars.
\subsection{General collapsing spacetime and equilibrium conditions}

The metric of a general spherically symmetric collapsing system can be written as
\begin{eqnarray}
ds^2 = - e^{2\nu(r,t)} dt^2 + {R'^2\over G(r,t)}dr^2 + R^2(r,t) d\Omega^2\, ,
\label{genmetric}
\end{eqnarray}
where $\nu(r,t)$, $R(r,t)$ and $G(r,t)$ are functions of local coordinates $r$ and
$t$.  Since none of the functions depends upon angular coordinates, the metric given above can be used to describe a spherical
gravitational collapse. Due to the presence of the undetermined functions(i.e., $\nu(r,t)$, $R(r,t)$ and $G(r,t)$) of $r$ and $t$, the above metric can describe any possible type of collapsing scenarios. For example, the collapsing matter can develop inhomogeneous energy density and pressure starting
from a homogeneous distribution of dust. There exists a scaling degree of freedom in the above-mentioned metric and using that freedom one can write \cite{JMN11}:
\begin{eqnarray}
R(r,t)= r f(r,t)\,,
\label{scaling}
\end{eqnarray}
where $f(r,t)$ is a scale factor that can depend upon both radial and temporal coordinates. One can set $f(r,t)$ as unity at the starting point of the collapse which implies $R(r,0)= r$. In the case of an isotropic fluid whose energy-momentum tensor components are given as
$T^0_{\,\,\,\,0}=-\rho$, $T^1_{\,\,\,\,1}=T^2_{\,\,\,\,2}=T^3_{\,\,\,\,3}=P$, the energy density and pressure  can be written as
\begin{eqnarray}
  \rho=\frac{F^\prime}{R^2 R^\prime}\,,\,\,\,\,
  P = -\frac{\dot{F}}{R^2 \dot{R}}\,,\,\,\,\,
\label{eineqns}
\end{eqnarray}
where $F(r,t)$ is the Misner-Sharp mass term and it is related to the total relativistic mass inside a particular comoving shell labeled by $r$ and $t$ \cite{MisnerSharp64, JMN11}. In this paper, a dot and a prime above any function are used to specify a derivative
of that function with respect to coordinate time and radius, respectively. Using the Eq.~(\ref{eineqns}) and the Einstein equations, one can write down the following mathematical expression for the Misner-Sharp mass term:
\begin{eqnarray}
F = R\left(1 - G + e^{-2\nu} \dot{R}^2\right)\,.
\label{ms}  
\end{eqnarray} 
In order to remove the non-diagonal terms of Einstein's equations we need  the following equation to be satisfied:
\begin{eqnarray}
\dot{G}=2\frac{\nu^\prime}{R^\prime}\dot{R}G\,.
\label{vanond}
\end{eqnarray}
Using the $G_2^2$ term and $\nabla_{\nu}T^{\mu\nu}=0$, one can get an equation which actually relates  $\rho, P, \nu^{\prime}$:
\begin{equation}
    \nu'=-\frac{P'}
{\rho+P} \,\, .
\label{pthecon}
\end{equation}
Along with the five Eqs.~(\ref{eineqns}, \ref{ms}, \ref{vanond}, \ref{pthecon}), in order to describe a physically viable collapse solution, the following conditions should always be satisfied \cite{JMN11}:
\begin{itemize}
\item Initially, the energy density and pressure should be regular at $r=0$ \cite{JMN11}. This implies that all of 
these quantities should be finite at the initial stage of collapse and the gradient of pressure should 
vanish at the center.
\item The collapsing matter should satisfy the weak energy condition throughout the collapse. As we know, the weak energy condition states that the energy density of matter with respect to any frame should be greater than zero i.e., $T_{\mu\nu}u^{\mu}u^{\nu}\geq 0$, where $u^{\mu}$ is the temporal orthonormal basis of a frame \cite{Poisson}. The above inequality implies: 
 \begin{eqnarray}
 \rho \geq 0, \, \, \rho+P > 0,\, \,.
 \end{eqnarray}
 \item A shell-crossing singularity may form inside a collapsing matter  when two different shells cross each other during the collapse ~\cite{Hellaby, Szekeres, Joshi:2012ak}, i.e., when $R^{\prime}=0$. In this paper, we avoid this type of singularity. To avoid this singularity, we
   require
\begin{eqnarray}
 R'(r,t) = f(r,t)+ r f'(r,t) > 0\, .
 \label{regcon}
\end{eqnarray}
\end{itemize} 
In the realm of general relativity, a self-gravitating system can reach an equilibrium state \cite{JMN11, Joshi:2013dva, Bhattacharya:2017chr, Dey:2019fja}. The conditions for the equilibrium state of a spherical over-dense region of physical radius $R$ is
\begin{eqnarray}
\lim_{t\to\mathcal{T}}\dot{R}=\lim_{t\to\mathcal{T}}\ddot{R}=0\,\, ,
\label{eqlb}
\end{eqnarray}
where the limit is taken for a large comoving time  $\mathcal{T}$. 

Using Eq.~(\ref{eqlb}) and Eq.~(\ref{scaling}), we can write down the following equilibrium condition
\begin{eqnarray}
\lim_{t\to\mathcal{T}}\dot{f}_e(r) = \lim_{t\to\mathcal{T}}\ddot{f}_e(r) =0\,,
\label{stabc}
\end{eqnarray}
where we use subscript $e$ to denote the equilibrium value of any
quantity as
$$f_e(r) \equiv \lim_{t \to \cal{T}} f(r,t)\,,$$
where $\cal{T}$ is very large time. Here, in
this section, we assume the existence of a final equilibrium
configuration of the gravitationally collapsing fluid. For the collapsing system that is purely integrable, we can analytically show the whole dynamics of the collapsing system starting from the initial state to the equilibrium configuration and we can also write down the equilibrated final space-time. In this section, we discuss the equilibrium condition of a general collapsing space-time defined by the metric in Eq.~(\ref{genmetric}). Therefore, for the general case, it is very difficult to realize the whole dynamics of the collapsing space-time.

In Sec.~(\ref{themodel}), we consider an integrable homogeneous system and analytically show how the system reaches the equilibrium configuration.   Using Eq.~(\ref{stabc}) and Eq.~(\ref{ms}), we can write down the following relation between $G(r,t)$ and $F(r,t)$ at the equilibrium state
\begin{equation}
G_e(r)=1-\frac{F_e(r)}{R_e}~,
\label{geq}
\end{equation}
where $G_e(r) = \lim_{t \to \cal{T}} G(r,t)\,,$ and $F_e(r) = \lim_{t \to \cal{T}} F(r,t)\,$.
We can write down the final form of the pressure at equilibrium ($P_e$) by using the equilibrium (static) Einstein equations, which can be 
easily obtained from the metric at equilibrium \cite{Joshi:2013dva}. The energy density and pressure of an isotropic fluid that is at the equilibrium state are as follows \cite{Joshi:2013dva}
\begin{eqnarray}
\rho_e &=& \frac{F_{,R_e}}{R_e^2}\,\, ,\\
P_e &=& \frac{2\nu_{e,R_e}}{R_e}G_e -\frac{F_e}{R_e^3}\,\, .
\label{eqbrhoP}
\end{eqnarray}
\subsection{The covariant form of the null result of the double derivative of the moment of inertia ($I$)}
One may relate the above-mentioned equilibrium conditions with the virialization condition using Eq.~(\ref{inertia1}).
The Newtonian virial theorem states that after a large enough time $$\Bigg\langle\frac{d^2I}{dt^2}\Bigg\rangle_{\tau\rightarrow\infty}=0.$$
One can write down a general relativistic covariant form of the above equation as,
\begin{equation}
\lim_{\tau \to \cal{T}}u^{\alpha}\nabla_{\alpha}u^{\beta}\nabla_{\beta}I=0\,\, ,
\label{vireq}
\end{equation}
where $u^{\alpha}$ is the timelike four-velocity of the comoving frame of a collapsing object, $\tau$ is the proper time in that comoving frame, $\cal{T}$ implies a large proper time limit and $I$ is the moment of inertia which is a scalar and it does not co-inside with the Newtonian expression of the moment of inertia in strong field regime. The general relativistic expression of the moment of inertia has been investigated in many papers \cite{Adams1,Adams2,Adams3}. For a spherically symmetric compact object with homogeneous density and constant equation of state (i.e., $\omega = \frac{P}{\rho}$), the moment of inertia can be written as \cite{Adams1}, $$I=\mathcal{C}F R^2$$ where $\mathcal{C}$ is a proportionality constant, $R$ is the physical radius and $F$ is the Misner-Sharp mass term. For the homogeneous scenario, the components of the four-velocity of the comoving frame of a collapsing object are $u^{\alpha}=\lbrace1,0,0,0\rbrace$.
Now, we can expand Eq.~(\ref{vireq}) as,
\begin{eqnarray}
    \lim_{\tau \to \cal{T}}u^{\alpha}\nabla_{\alpha}u^{\beta}\nabla_{\beta}I&=& \lim_{\tau \to \cal{T}}\left(a^{\beta}\partial_{\beta}I+u^{\alpha}u^{\beta}\nabla_{\alpha}\partial_{\beta}I\right)\nonumber \\
    &=& \lim_{\tau \to \cal{T}}\partial^2_0 I = 0
\end{eqnarray}
where $a^{\beta}=u^{\alpha}\nabla_{\alpha}u^{\beta}$. Using the expression for the moment of inertia and the Eq.~(\ref{vireq}), we get
\begin{eqnarray}
 \lim_{\tau \to \cal{T}}\left[\Ddot{F}R^2+2F\left(\Dot{R}^2+R \Ddot{R}\right)\right] = 0\,\, .   
\end{eqnarray}
Using Eqs.~(\ref{ms}, \ref{vanond}) and the regularity conditions one can show that the above equation implies
\begin{eqnarray}
    \lim_{\tau \to \cal{T}}\Dot{R}=\lim_{\tau \to \cal{T}}\Ddot{R}=0\,\, .
\end{eqnarray}
Therefore, when the Eq.~(\ref{vireq}) is satisfied, a collapsing, spherically symmetric, homogeneous compact object should follow the equilibrium conditions.
Hence, we show that for a homogeneous scenario, the above equilibrium condition satisfies the Eq.~(\ref{vireq}); however, we do not know whether the Eq.~(\ref{vireq}) is the general relativistic condition of virialization. Further investigation is needed here.

In the next subsection, we redefine the equilibrium condition in an invariant way. We write down the conditions using Cartan scalars.
	\subsection{Equilibrium Conditions Defined Invariantly}

In order to investigate the spinor structure \cite{penrose1} of the general collapsing space-time (Eq.~(\ref{genmetric})), first we need to construct a fixed null coframe of that space-time. The null coframe of the general collapsing space-time can be constructed as
	\begin{equation}
	\begin{aligned}
			\ell &= \frac{1}{\sqrt{2}}{(t^0 - t^1)}= \frac{1}{\sqrt{2}}\left( e^{2\nu(r,t)}dt - \frac{R^{\prime}}{\sqrt{G(r,t)}} dr\right), \\ 
			 n &=  \frac{(t^0 + t^1)}{\sqrt{2}} =\frac{1}{\sqrt{2}}\left( e^{2\nu(r,t)}dt +\frac{R^{\prime}}{\sqrt{G(r,t)}} dr\right), \\
			m &= \frac{(t^2 - it^3)}{\sqrt{2}} =  \frac{1}{\sqrt{2}}\left(Rd\theta  - iR\sin(\theta) d\phi \right),\\ 
			\bar{m} &= \frac{(t^2 + it^3)}{\sqrt{2}} = \frac{1}{\sqrt{2}}\left(Rd\theta  + iR\sin(\theta) d\phi\right),
	\end{aligned}
	\end{equation}
	
	\noindent
where $t^0, t^1, t^2$ and $t^3$ are the bases of the orthonormal frame at a point in the Manifold $\mathcal{M}$.
Using the null frame bases, one can define the bases of the spin frame $(o^A, i^A)$ and its conjugate spin frame $(o^{A^{\prime}}, i^{A^{\prime}})$ in the following way: $l^a =o^A o^{A^{\prime}},~ n^a = i^A i^{A^{\prime}},~ m^a = o^A i^{A^{\prime}},~ \bar{m}^a=i^A o^{A^{\prime}}$ \cite{penrose1}.
 This spin frame will be used along with the Newman-Penrose formalism to employ the Cartan-
Karlhede algorithm and classify the spacetime. The non-zero spin coefficients ($\gamma_{\boldsymbol{AB^{\prime}CD}}$) of the corresponding spin frame of the above-mentioned null frame are:
\begin{eqnarray}
    \gamma_{\boldsymbol{01^{\prime}11}}=\mu&=&\frac{G(r,t)+\dot{R}e^{-2\nu(r,t)}}{\sqrt2 R},\\
\gamma_{\boldsymbol{10^{\prime}00}}=    \tilde{\rho}&=&\frac{G(r,t)-\dot{R}e^{-2\nu(r,t)}}{\sqrt2 R}\, ,\\
 \gamma_{\boldsymbol{10^{\prime}10}}= \alpha&=&-\beta=-\frac{\sqrt2 \cot(\theta)}{4 R}\, ,\\
\gamma_{\boldsymbol{11^{\prime}10}}=\gamma &=&\frac{-4\nu^{\prime}e^{2\nu}G^{\frac32}-2G\dot{R}^{\prime}+R^{\prime}\dot{G}}{4\sqrt2 G R^{\prime}e^{2\nu}}\,\, ,\\
\gamma_{\boldsymbol{00^{\prime}10}}=\epsilon &=&\frac{-4\nu^{\prime}e^{2\nu}G^{\frac32}+2G\dot{R}^{\prime}-R^{\prime}\dot{G}}{4\sqrt2 G R^{\prime}e^{2\nu}}.
\end{eqnarray}

\noindent
	Note that the above spin coefficient $\tilde{\rho}$ is distinct from the energy-density fluid quantity $\rho$.We also define the frame operators $D\equiv \ell^a \nabla_a\equiv o^A o^{A^{\prime}} \nabla_{A A^{\prime}}$, $\Delta \equiv n^a \nabla_a \equiv i^A i^{A^{\prime}} \nabla_{A A^{\prime}} $, and note that the combined operator has the following property
	
	\begin{equation}
		D - \Delta \propto \partial_t.
	\end{equation}

	\subsubsection{Brief review of the Cartan-Karlhede Algorithm}

The method for determining the local equivalence of space-times involves the definition of an invariant frame for a space-time. The Cartan scalars are then the set of scalars that are the result of contractions of the curvature tensors and the invariant frame vectors. Note that by invariant frame, we mean that the frame is defined to be invariant up to an isotropy group. The derivation of these Cartan scalars is done through the Cartan-Karlhede algorithm \cite{cartan-karlhede-alg}. We briefly list the Cartan Karlhede algorithm below, and then employ the algorithm for the spherically symmetric perfect fluid space-time discussed so far.

\noindent
The Cartan Karlhede Algorithm is summarized as in \cite{kramer}:

1. Set the order of differentiation, $q$, to zero.

2. Calculate the derivatives of the Riemann tensor up to the $q$th order.

3. Find the canonical form of the Riemann tensor and derivatives.

4. Using Lorentz transformations, put the Riemann tensor and its derivatives in canonical form, and note the number of transformations that keep the Riemann tensor and its derivatives invariant (the linear isotropy group is this group of transformations).

5. Determine the number $t_q$ of independent functions of the coordinates in the Riemann tensor and its derivatives, in the canonical form.

6. If the isotropy group and the number of independent functions are the same as in the previous step, let $p+1=q$, and stop. Else, increase $q$ by $1$, and continue from step 2.

\noindent
The set of independent functions determined by the algorithm are the Cartan scalars for this particular invariant frame. We also note that any scalar constructed from the Cartan scalars is referred to as an extended Cartan scalar.

	At zeroth order, the Weyl and Ricci spinors each only contain one non-zero component which using the field equations can be written in terms of the above metric functions and fluid quantities as \cite{Coley:2017woz, Layden:2022oxu}:
	\begin{equation} 
		\begin{aligned}
			\Psi_2 = \frac{\rho}{6}  - \frac{F}{2R^3}, \ \Phi_{00} &= \Phi_{22} = 2\Phi_{11} = \frac{(\rho + p)}{4},
		\end{aligned}
  \label{psiphi}
	\end{equation}

	\noindent
 where $\Psi_2 =C_{abcd}l^a m^b\bar{m}^c n^d = C_{1342}$, $\Phi_{00}=\frac12 R_{ab}l^al^b$, $\Phi_{22}=\frac12 R_{ab}n^an^b$ and $\Phi_{11}=\frac14 R_{ab}(l^an^b+m^a\bar{m}^b)$, where $C_{abcd}$ and $R_{ab}$ are the Weyl and Ricci tensors, respectively.
	At first order, the derivatives of the Weyl tensor only contains boost weight +1 and -1 components, here given in terms of derivatives of the zeroth order scalars and spin coefficients, using the Bianchi identities. The boost weight +1 components are
	
	\begin{eqnarray}
	    		C_{1432;1}&=&C_{1342;1} = \frac12C_{1212;1} = \frac12C_{3434;1} = D\Psi_2, \nonumber\\
	\end{eqnarray}
	\begin{equation}
		C_{1214;3} = C_{1213;4} = C_{1343;4} = C_{1434;3} = 3\tilde{\rho}\Psi_2,
	\end{equation}
	
	\noindent
	and the boost weight -1 components are
	\begin{eqnarray}
		C_{1432;2} &=& C_{1342;2}=\frac12C_{1212;2}=\frac12C_{3434;2} =\Delta \Psi_2, \nonumber \\
	\end{eqnarray}
	
	\begin{equation}
		C_{1223;4} = C_{1224;3} = C_{2334;4} = C_{2443;3} = 3\mu\Psi_2.
	\end{equation}

	\noindent
	In an invariant frame, the frame derivatives $D\Psi_2$ and $\Delta \Psi_2$ are Cartan scalars, and can be combined to construct a new scalar that is proportional to the time derivative of the weyl scalar $\Psi_2$ as follows:
	
	\begin{equation}
		\mathcal{C} \equiv D\Psi_2 - \Delta \Psi_2 \propto \partial_t \Psi_2.
	\end{equation}
	
	\noindent
Using the expression for $\Psi_2$ in Eq.~(\ref{psiphi}), the expression for $\partial_t \Psi_2$ can be written as
\begin{eqnarray}
    \partial_t \Psi_2 = \frac{\dot{\rho}}{6}-\frac{\dot{F}}{2R^3}+\frac{3F}{2R^4}\dot{R}.
    \label{delpsi2}
\end{eqnarray}
	At the equilibrium state, when  $\lim_{t\to\mathcal{T}}\dot{R}=\lim_{t\to\mathcal{T}}\ddot{R}=0$ (Eq.~(\ref{eqlb})), $\lim_{t \to \cal{T}} \rho(r,t)=\rho_e(r),$ (Eq.~(\ref{eqbrhoP})) and $\lim_{t \to \cal{T}} F(r,t)=F_e(r)$. Therefore, we can write 
 \begin{equation}
     \lim_{t \to \cal{T}} \partial_t \Psi_2 = \lim_{t \to \cal{T}} \mathcal{C} =0
 \end{equation}
	or equivalently, one can write
	
	\begin{equation}
	\lim_{t \to \cal{T}} 	D\Psi_2 =\lim_{t \to \cal{T}}  \Delta \Psi_2.
	\end{equation}
Now, from the expression of energy density ($\rho$), Misner-Sharp mass ($F$) and $\dot{G}$  in Eq.~(\ref{eineqns}), Eq.~(\ref{ms}) and Eq.~(\ref{vanond}), respectively, it can be shown that $\dot{\rho}=0$ and $\dot{F}=0$ do not necessarily imply $\ddot{R}=0$. The null values of $\dot{\rho}$ and $\dot{F}$ are always obtained when $\dot{R}=0$, where $\ddot{R}$ may or may not be equal to zero. Therefore, $D\Psi_2 = \Delta \Psi_2$ implies any one of the following states: an equilibrium state, a bouncing state (i.e., when $\dot{R}=0, \ddot{R}>0$), a turnaround state (i.e., when $\dot{R}=0, \ddot{R}<0$). Therefore, we need to find whether any other combinations of Cartan scalars uniquely define an equilibrium state. Since $D\Psi_2$ and $\Delta \Psi_2$ are Cartan scalars, $D^2 \Psi_2$, $\Delta^2 \Psi_2$ are also Cartan scalars, additionally, the spin coefficients $\epsilon,\ \gamma$ are extended Cartan scalars, as they can be expressed in terms of other Cartan scalars. Therefore, their combination $(D-\Delta)^2\Psi_2 = D^2\Psi_2 + \Delta^2\Psi_2 -D\Delta \Psi_2 - \Delta D \Psi_2$ is also a scalar and that is proportional to $\partial_t^2 \Psi_2$. Now the expression for $\partial_t^2 \Psi_2$ is:
\begin{eqnarray}
    \partial_t^2 \Psi_2 = \frac{\ddot{\rho}}{6}-\frac{\ddot{F}}{2R^3}+\frac{3\dot{F}}{R^4}\dot{R}-\frac{6F}{R^5}\dot{R}^2+\frac{3F}{2R^4}\ddot{R},
\end{eqnarray}
which shows $\partial_t^2 \Psi_2 =0$ or $(D-\Delta)^2\Psi_2 = 0~ \forall r$ uniquely implies $\dot{R}=\ddot{R}=0$. Therefore, finally, we can define the equilibrium condition of gravitational collapse by using Cartan scalars as:
\begin{eqnarray}
    \lim_{t \to \cal{T}}(D-\Delta)^2\Psi_2 = 0~ \forall r.
\end{eqnarray}

The discussion of equilibrium states can additionally be formulated in an invariant way by constructing scalar polynomial invariants out of the derivatives of the curvature tensor. This way of determining properties of spacetimes using scalar polynomial invariants has been done for other spacetimes, and in particular has been used for identifying black hole horizons in exact solutions \cite{mcnuttpage_spi}. Note that the squared and cubed Weyl scalar curvature invariants satisfy the Type-D condition $I^3 - 27J^2 = 0$:
\begin{equation}
    I = 3\Psi_2^{\ 2}, \ J=\Psi_2^{\ 3}.
\end{equation}

Additionally, for example, we also have the following invariant expressions
\begin{eqnarray}
     I_{,a}I^{,a} \propto \Psi_{2}^{\ 2} D\Psi_2 \Delta \Psi_2, \\ C_{abcd;ef}C^{abcd;ef} \propto (D - \Delta)^2\Psi_{2} + f(C_i), \\ I_{;a}^{\ ;a} \propto \Psi_{2} \Delta D \Psi_2 + g(C_i),
\end{eqnarray}

 \noindent
 where $f(C_i),g(C_i)$ are both polynomial functions of the Cartan scalars, including spin coefficients (which are extended Cartan scalars). In practice, constructing such a scalar polynomial invariant to detect the equilibrium condition is possible, but rather non-trivial, so we present these invariants here as an indicator of how the method could also be applied using scalar curvature invariants.

\section{Review on Top-Hat collapse model and its possible Modification}
\label{tophat}
In this section, we first briefly review the top-hat collapse model. Then we discuss works in which the model is modified by considering the dark energy effect in the over-dense region of dark matter. 
\subsection{Top-Hat Collapse model}
As alluded to before, in the top-hat collapse model the dynamics of the over-dense region of dark matter are described by the closed FLRW metric (Eq.~(\ref{FLRWMetric})). 
Using the Einstein field equation for the closed FLRW metric, one can write down the governing Friedmann equation as
\begin{eqnarray} 
\frac{H^2}{H_0^2}=\Omega_{m0}\left(\frac{a_0}{a}\right)^3 + (1-\Omega_{m0})\left(\frac{a_0}{a}\right)^2 \, ,
\label{fried}
\end{eqnarray}
where $H=\dot{a}/a$ is the Hubble parameter for the over-dense
region and $H_0,\,a_0$ are the values of $H$ and $a$ at the time when the influence of the background cosmological expansion on the dynamics of over-dense regions becomes negligible. After that time, the over-dense regions start behaving like sub-universes which have their own distinguishable dynamics. In the above Friedmann equation $\Omega_{m0}=\rho_0/\rho_{c0}$, where $\rho_{c0} =
3H_0^2$, and $\rho_0$ is the matter density when $a=a_0$. In
the standard top-hat collapse model, the dark-energy effect is not considered, and therefore $\Lambda=0$.
The parametric solution of Eq.~(\ref{fried}) can be written as \cite{GunnGott72}:
\begin{eqnarray}
 a &=& {a_0 \Omega_{m0}\over 2 (\Omega_{m0}-1)} (1- \cos\theta)\,\, \nonumber\\
 t &=& {\Omega_{m0}\over 2 H_0 (\Omega_{m0}-1)^{3/2}} (\theta-\sin\theta)\, .
\label{at}
\end{eqnarray}
For the over-density of the sub-universe, we assume $\Omega_{m0}>1$.  From the above solution of the Friedmann equation, it can be understood that the spherical over-dense region expands
first and after a time period, when $\theta=\pi$, the physical radius of the over-dense region reaches its maximum value $R_{max}$. Using Eq.~(\ref{at}), one can obtain the maximum value of the scale factor and the corresponding time:  
$$a_{\rm max}={a_0 \Omega_{m0}\over (\Omega_{m0}-1)}\,,\,\,\,\, t_{\rm
  max}={\pi \Omega_{m0}\over 2 H_0 (\Omega_{m0}-1)^{3/2}}\,.$$
At $t=t_{max}$, one can calculate the ratio of the density of the over-dense region and the density of the background ($\bar{\rho}$):
\begin{eqnarray}
  \frac{\rho(t_{\rm max})}{\bar{\rho}(t_{\rm max})} = \frac{\Omega_{m0}\rho_{c0} \left(\frac{a_0}{a}\right)^3}
       {\bar{\rho}_{c0} \left(\frac{\bar{a}_0}{\bar{a}}\right)^3}\,.
\label{ratio1}
\end{eqnarray}
From now on, all the background parameters and variables will be denoted by an over-bar.
We can assume $a_0 \sim \bar{a}_0$ and $\rho_{c0} \sim \bar{\rho}_{c0}$ at the time when the over-dense regions detach from the background and with this assumption we can write down the above ratio as:
\begin{eqnarray}
  \frac{\rho(t_{\rm max})}{\bar{\rho}(t_{\rm max})} = \Omega_{m0}\left[\frac{\bar{a}(t_{\rm max})}
    {a_{\rm max}}\right]^3\,.
\label{ratio1}
\end{eqnarray}
Generally, flat FLRW spacetime is used to describe the dynamics of the background universe. During the matter-dominated era, the scale factor of the background universe can be written as: 
$\bar{a}(t)=\bar{a}_0 \left(\frac{3}{2}\bar{H}_0 t\right)^{2/3}$.
Using the expression of $\bar{a}(t)$ 
in Eq.~(\ref{ratio1}) and considering the assumption $\bar{H}_0\sim H_0$, we get
\begin{eqnarray}
\frac{\rho(t_{\rm max})}{\bar{\rho}(t_{\rm max})} = \frac{9\pi^2}{16} \sim 5.55\,, 
\label{impfac}
\end{eqnarray}
which implies that after the primary expansion phase, the spherical over-dense region starts collapsing when the density of the over-dense region is $5.55$ times the density of the
background universe. The solution of Eq.~(\ref{fried}) implies that
the final state of collapsing matter is a singularity when $\theta= 2\pi$ and that singularity is a space-like singularity i.e., the singularity is covered by trapped surfaces. Therefore, for this type of collapsing scenario (i.e., for spherically symmetric, homogeneous dust collapse), a black hole should always form as a final state of collapse. To describe a stable configuration as a final state of a spherically symmetric, homogeneous dust collapse, one needs to introduce some mechanism that can stabilize the collapsing matter before the formation of the black hole. As discussed before, in the top-hat model, the Newtonian virialization technique is used to stabilize the collapsing system.

As discussed in the introduction, a self-gravitating collapsing system virializes when
$\langle T\rangle=-\frac12 \langle V_T \rangle\, ,$
\begin{figure*}
\includegraphics[width=185mm]{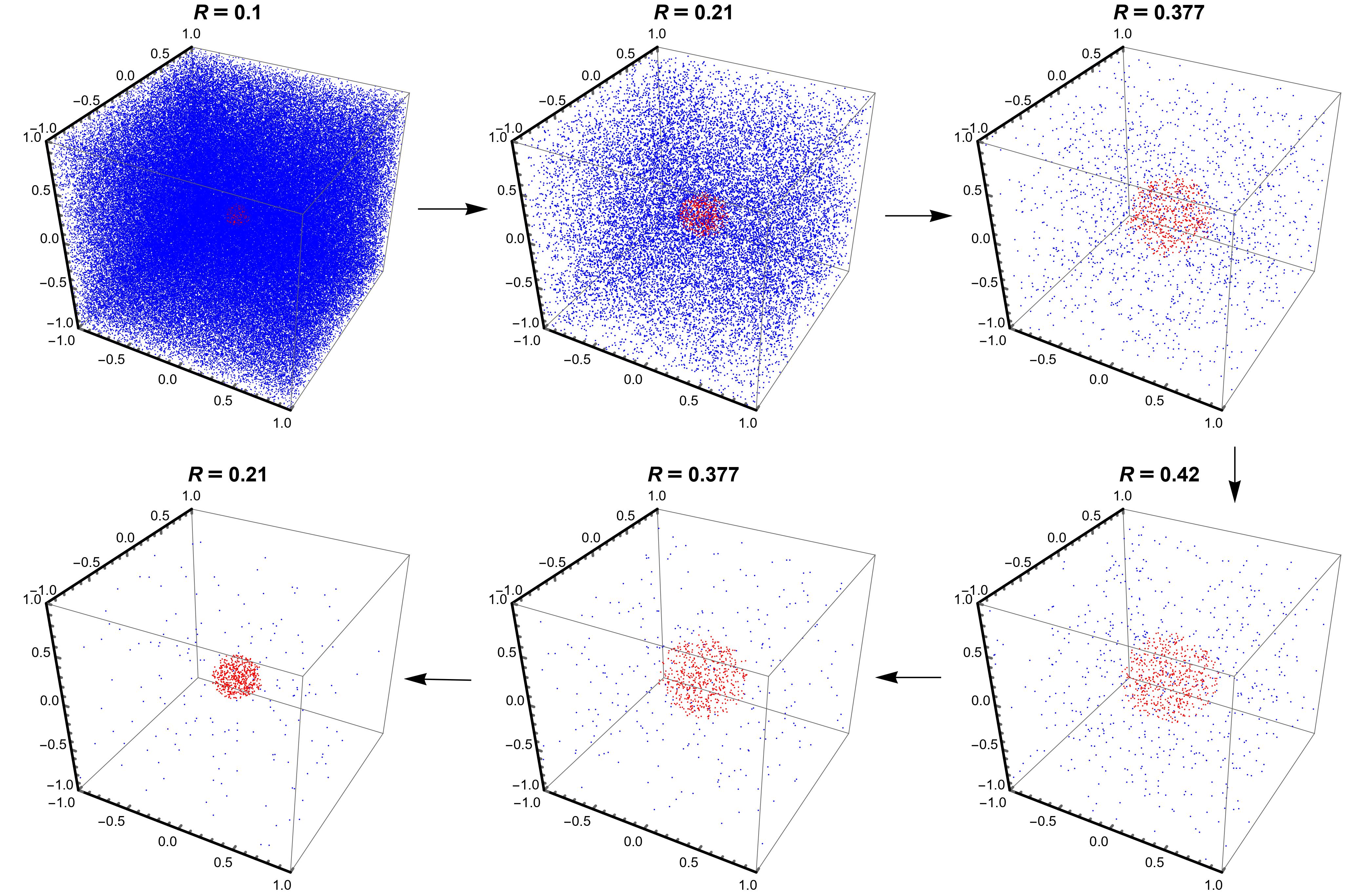}
 \caption{The figure shows the evolution of the over-dense region in the dark matter field, where the over-dense regions are shown by the spherically symmetric regions which are filled with red points, and the background is shown by the region with blue points. The number of points per unit volume in both regions describes the density of matter in those regions. It can be seen how the over-dense region expands first with the background (from the physical radius $R=0.1$ to $R=0.377$ ) and after the turn-around point (with physical radius $R=0.42$), how it collapses until the virialization state (with physical radius $a=0.21$) is achieved. The arrows in the figure indicate the direction of the evolution. The total number of points inside the spherical over-density is $10^3$, which is considered to be always conserved. This simulation is done, considering the Top-Hat collapse model.  }
 \label{tophatevol}
\end{figure*}
Now, when a spherical body having mass $M$ reaches its maximum physical radius ($R_{max}$) at the turnaround time ($t=t_{max}$), momentarily it's kinetic energy becomes zero. Since the virialization process discussed above is a Newtonian process, here the total mass $M$ is the non-relativistic mass of the spherical body. At the turnaround time ($t=t_{max}$) all the energy of the spherical body is the potential energy which can be written as:
\begin{equation}
V_T=-\frac{3M^2}{5R_{max}}\, .
\end{equation}
One can show that at the physical radius $R=\frac{R_{max}}{2}$, 
the kinetic energy of the spherical body becomes half of the absolute value of the potential energy:
\begin{eqnarray}
V_T&=& -\frac{6M^2}{5R_{max}}\, ,\nonumber\\
T&=& \frac{3M^2}{5R_{max}} = -\frac{V_T}{2}\, .
\end{eqnarray}
Therefore, according to the Newtonian virial theorem, the spherical body virializes when the physical radius $R=\frac{R_{max}}{2}$ or the scale factor $a =\frac12 a_{max}$.
At this point, it is considered that the collapsing system undergoes violent relaxation \cite{Lynden-Bell67} processes and achieves a static, virialized configuration.
From  Eq.~(\ref{at}), one can show that the over-dense region reaches the virialized configuration when $\theta = \frac{3\pi}{2}$. The virialization time $t_{\rm vir}=t_{\rm max}\left(\frac32+\frac{1}{\pi}\right)\sim1.81t_{\rm max}$. Therefore, according to the top-hat collapse model, the over-dense region expands first, and at time $t_{\rm max}$, it starts collapsing due to its own gravity, and finally, the system virializes to a stable structure at the time $t_{\rm vir}$. At $t=t_{\rm vir}$ one can write the ratio as:
\begin{eqnarray}
  \frac{\rho(t_{\rm vir})}{\bar{\rho}(t_{\rm vir})} \sim 145\,.
\label{ratio2}
\end{eqnarray}
There are various models where it is shown that the density of the collapsing spherical over-dense region can vary
between 145-200 times the density of the background universe. In those models, extrapolation of the final time up to the time of the collapse is considered. There is another way to define density contrast. In that definition, a spherical body with an initial physical radius ($R_i$) is considered, where $R_i$ is proportional to the initial scale factor $a_i$ (i.e., some value scale factor during the initial phase of expansion) which is not equal to $a_0$.
At that initial time, the density of the spherical body is equal to the background density $\bar{\rho}$. After some time, we can write the physical radius of the spherical body  $R \propto (a_i+\Delta a)$, where $\Delta a$ is the small change of the scale factor. The density of the spherical body also becomes $\rho=\bar{\rho} + \Delta \rho$. If we assume zero flux of matter across the boundary of the over-dense region during the small change of physical radius, then we can define a linear density contract $\delta$ as
$$\delta \equiv \frac{\Delta \rho}{\bar{\rho}}=-3\frac{\Delta
  a}{a_i}\,.$$
Now if we do Taylor's expansion of Eq.~(\ref{at}) as a series
in $\theta$ and use the first few terms, we can write down the following value of linear density contrast at the virialization time $t_{vir}$:
\begin{eqnarray}
\delta_{\rm vir} \sim 1.69\,.
\label{dtc}
\end{eqnarray}
With this approximation, one can calculate the density ratio between the density of the over-dense region and the density of the background at virialization:
\begin{eqnarray}
  \frac{\rho(t_{\rm vir})}{\bar{\rho}(t_{\rm vir})} \sim
  170 - 200\,.
\label{ratio3}
\end{eqnarray}

In fig.~(\ref{tophatevol}), considering the top-hat collapse model, we show how the spherical over-density of dark matter evolves with time. In that figure, we consider a spatially flat FLRW background which is dominated by dark matter. At the starting point of the evolution, when the over-density detaches from the background expansion, $\Omega_{m0}=1.315$. As we know, in the top-hat collapse model, the over-dense regions are modelled by closed FLRW spacetime. Therefore, in the simulation shown in fig.~(\ref{tophatevol}), the total number of points ($10^3$) inside the over-dense region is always fixed. In that figure, the over-dense regions are shown by the spherically symmetric regions which are filled with red points, and the background is shown by the region with blue points. It also can be seen that the over-dense region reaches the virialized configuration when $R_{vir}=\frac{R_{max}}{2}$.

 As it has been discussed in the introduction, there is literature in which the effect of dark energy on the virialization of dark matter is investigated \cite{Lahav:1991wc, Steinh, Shapiro, Basilakos:2003bi, Caldwell:2003vq, Horellou:2005qc, Maor:2005hq, Percival:2005vm, mota2006, Wang:2005ad, Basilakos:2006us, Maor:2006rh, Basilakos:2009mz, Basilakos:2010rs, Leewang, Chang:2017vhs}. In the next subsection, we briefly review how the inclusion of dark energy effect modifies the top-hat collapse model. 
\subsection{Possible modifications of Top-Hat collapse model}
\label{virwithdark}
If we consider a combination of dark matter (DM) and dark energy (DE) as a two fluid system inside the over-dense region, then the total gravitational potential energy can be written as \cite{Maor:2005hq}
\begin{eqnarray}
V_T = \frac12 \int_v &\rho_{DM}&\phi_{DM} ~dv + \frac12\int_v \rho_{DM}~\phi_{DE} ~dv \nonumber\\ &+& \frac12\int_v \rho_{DE}~\phi_{DM} ~dv+ \frac12\int_v \rho_{DE}~\phi_{DE}~ dv\,\, ,\nonumber\\
\label{VT1}
\end{eqnarray}
where $\rho_{DM}$ and $\rho_{DE}$ are the energy density of dark matter and dark energy, respectively, and $\phi_{DM}$ and $\phi_{DE}$ are the gravitational potential of dark matter and dark energy, respectively. The integration is done over the whole volume ($v$) of the spherical over-dense region. According to the non-zero values of the four integration written above, we can classify the two fluid system into the following four distinguishable scenarios,
\begin{itemize}
\item In the first scenario, the spherical over-densities of dark matter virializes, where the dark energy effect is totally negligible \cite{GunnGott72}. 
\item In the second scenario, there exists a non-negligible effect of the homogeneous dark energy which modifies the virialization process of the spherically symmetric over-dense regions of dark matter. In this scenario, though there exists a non-negligible effect of dark energy in the virialization process of dark matter, dark energy cannot cluster and virialize with dark matter \cite{Lahav:1991wc, Steinh, Shapiro, Horellou:2005qc}.
\item In the third scenario, dark energy clusters inside the over-dense regions of dark matter. However, dark energy does not virialize with dark matter. This scenario is known as the clustered dark energy scenario \cite{Basilakos:2003bi, Maor:2005hq, Basilakos:2006us, Basilakos:2009mz, Chang:2017vhs}, where
it is considered that from the beginning of the matter-dominated era, the dark energy moves synchronously with the dark matter on both the Hubble scale and the galaxy cluster scale.
\item In the fourth scenario, dark energy clusters and also virializes with dark matter inside the spherical over-dense regions \cite{Maor:2005hq}.
\end{itemize}

If we totally neglect the dark energy effect, then only the first integration would contribute  corresponding to the self-gravitating dark matter. This scenario is described by the top-hat collapse model, where one self-gravitating fluid, inside a spherical over-dense region, virializes \cite{GunnGott72}. The total potential energy inside the spherical over-dense region of radius $R$ can be written as
\begin{eqnarray}
V_T = \frac12\int_v &\rho_{DM}&\phi_{DM} ~dv = -\frac{3M^2}{5R}\, . 
\end{eqnarray} 
We know that with the above total potential energy, the spherically symmetric over-densities virialize when $\eta= \frac{R_{vir}}{R_{max}}=0.5$. 
\begin{figure}[h!]
\centering
{\includegraphics[width=90mm]{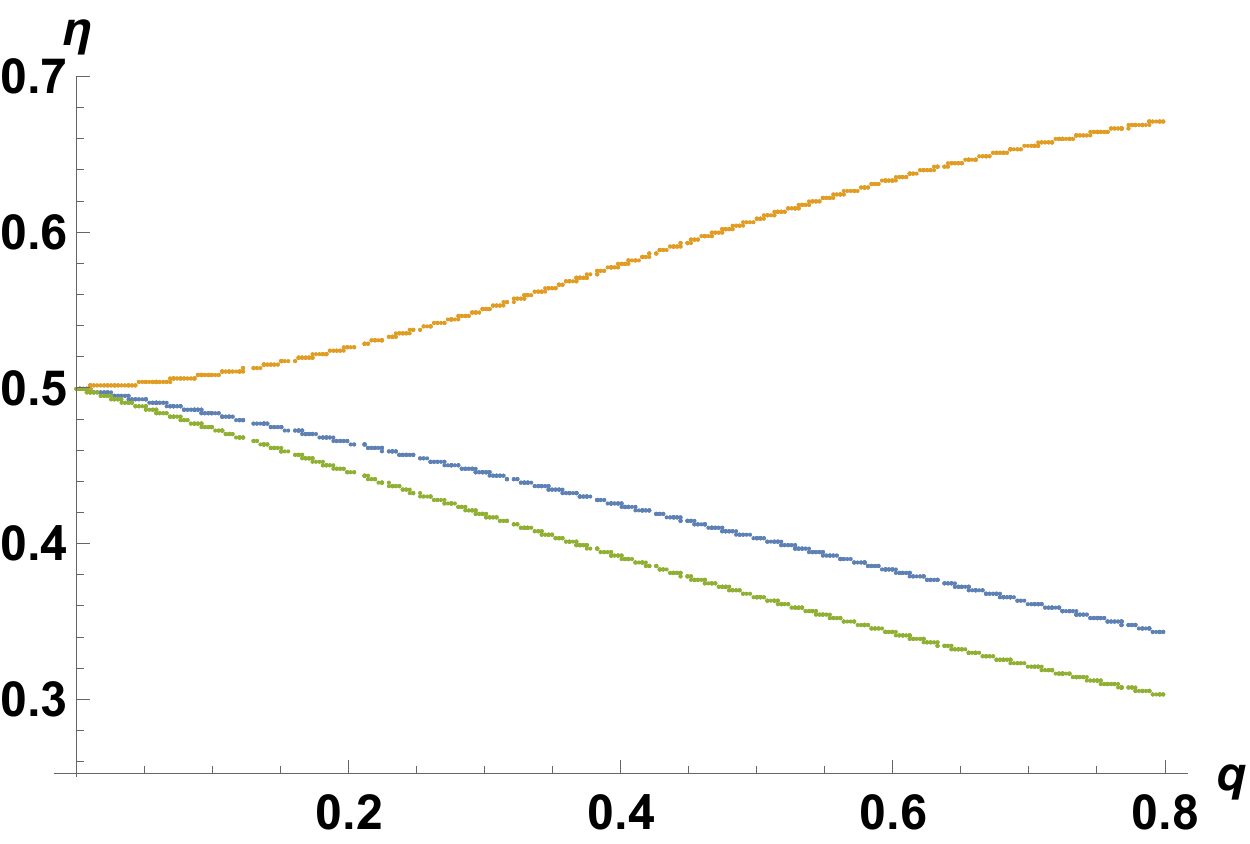}}
\caption{Figure shows how $\eta$ varies with $q$, where $\eta=\frac{R_{vir}}{R_{max}}$  $q=\left(\frac{\rho_{DE}}{\rho_{DM}}\right)_{a=a_{max}}$. The brown line depicts the behavior of $\eta$ in the case where dark energy with $\omega =-0.75$ clusters and virializes inside the spherical over-densities of dark matter and the green line shows the behavior of $\eta$ in the case where clustered dark energy with the same equation of state (i.e., $\omega=-0.75$) cannot virialize with dark matter. The blue line shows how $\eta $ changes with $q$ for homogeneous dark energy with $\omega=-1$ (i.e., a cosmological constant $\Lambda$).}
\label{etavsq1}
\end{figure}

In \cite{Lahav:1991wc} and \cite{Steinh}, the authors investigated the homogeneous dark energy effect on the virialization of the spherical over-densities, where in \cite{Lahav:1991wc}, the authors consider the homogeneous $\Lambda$ dark energy model and in \cite{Steinh}, the authors consider the homogeneous quintessence  dark energy model. In these scenarios, though the dark energy does not cluster and virialize inside the spherical over-densities of dark matter, the non-zero energy density and negative pressure of dark energy modify the virialization process of the over-densities which can be realized by the different values of $\eta$. For the homogeneous dark energy scenario, the total potential energy of the over-dense region can be written as \cite{Maor:2005hq}
\begin{eqnarray}
V_T = \int_v &\rho_{DM}&\phi_{DM} ~dv + \int_v \rho_{DM}~\phi_{DE} ~dv\,\, ,
\end{eqnarray}
which gives
\begin{eqnarray}
V_T= -\frac{3M^2}{5R} + \frac{3M^2}{10R} ~q(1+3\omega)\left(\frac{\bar{a}}{\bar{a}_{max}}\right)^{-3(1+\omega)}\left(\frac{R}{R_{max}}\right)^3 ,\nonumber\\
\label{VT2}
\end{eqnarray}
where $q=\left(\frac{\rho_{DE}}{\rho_{DM}}\right)_{a=a_{max}}$, which is the ratio of energy densities of dark energy and dark matter inside the spherical over-dense regions at the turn around time $t= t_{max}$ and $\omega$ is the equation of state of dark energy. As was mentioned, $\bar{a}, a$ represent the scale factor of background and spherical over-dense regions, respectively. At the time $t=t_{max}$, when the over-dense region reaches its maximum physical radius ($R_{max}$), the scale factor of the background is $\bar{a}_{max}$.  Using Eq.~(\ref{VT2}) and the virialization condition $\left(V_T + \frac12 R \frac{\partial V_T}{\partial R}\right)_{a=a_{vir}}=(V_T)_{a=a_{max}}$, we can get following cubic equation of $\eta$
\begin{eqnarray}
4Q\eta^3 \left(\frac{\bar{a}_{vir}}{\bar{a}_{max}}\right)^{-3(1+\omega)} -2 \eta (1+Q) +1 =0\,\, ,
\end{eqnarray}
where $Q=-(1+3\omega)\frac{q}{2}$.
One can verify from the above equation that for the vanishing value of $q$ (i.e., no dark energy effect), we get $\eta=0.5$, which we obtained earlier for the top-hat collapse model. For $\omega =-1$ ($\Lambda$ dark energy), one can get the following cubic equation for $\eta$ \cite{Lahav:1991wc, Shapiro}
\begin{eqnarray}
4q\eta^3  -2 \eta (1+q) +1 =0\,\, .
\end{eqnarray}  
Using the above equation and considering small value of $q$, the solution of $\eta$ can be written as \cite{Shapiro}
\begin{equation}
\eta= 0.5 - 0.25 q -0.125 q^2+ \mathcal{O}(q^3)\,\, .
\end{equation}
The above approximate solution of $\eta$ shows that due to the effect of the cosmological constant ($\Lambda$), the value of $\eta$ becomes less than $0.5$, which implies that the spherical over-densities need to collapse more to reach the virialization state. 

After the spherical over-densities reach the virialization state, the energy density of the homogeneous dark energy (with $\omega\neq -1$) inside the over-dense regions varies with time. As the universe expands, the energy density of the dark energy ($\omega\neq -1$) inside the virialized spherical regions decreases. This phenomenon is discussed elaborately in \cite{Maor:2005hq}. To avoid this problem, clustered and virialized dark energy model is proposed, where at the galaxy cluster scale, dark energy can cluster and virialize inside the over-dense regions of dark matter. In this case, the total potential energy of the spherical regions can be written as \cite{Maor:2005hq}
\begin{widetext}
\begin{eqnarray}
V_T = -\frac{3M^2}{5R}-(2+3\omega)\frac{3M^2}{5R}q\left(\frac{R}{R_{max}}\right)^{-3\omega}-(1+3\omega)\frac{3M^2}{5R}q^2\left(\frac{R}{R_{max}}\right)^{-6\omega}\,\,  ,
\end{eqnarray}
where each of the integrations in Eq.~(\ref{VT1}) has non-zero value. Using the above expression of total potential energy and the virialization condition, we get the following equation for $\eta$ \cite{Maor:2005hq}
\begin{eqnarray}
\left[1+(2+3\omega)q+(1+3\omega)q^2\right]\eta-\frac12(2+3\omega)(1-3\omega)q\eta^{-3\omega}-\frac12(1-6\omega)(1+3\omega)q^2\eta^{-6\omega}=\frac12\,\, .
\end{eqnarray} 
\end{widetext} 
There exists another scenario where the dark energy only can cluster inside the spherical over-densities; however, it cannot virialize at that scale. For this scenario, the total potential energy of the spherical regions can be written as
\begin{eqnarray}
V_T=-\frac{3M^2}{5R}\left[1+q\left(\frac{R}{R_{max}}\right)^{-3\omega}\right]\,\, ,
\end{eqnarray}
from which we get the following equation for $\eta$ \cite{Maor:2005hq}
\begin{eqnarray}
\eta(1+q)-\frac{q}{2}(1-3\omega)\eta^{-3\omega}=\frac12\,\, .
\end{eqnarray} 
In fig.~(\ref{etavsq1}), by the brown line, we show how $\eta$ changes with $q$ when dark energy can cluster and virialize inside the over-dense regions of dark matter. We can see that the value of $\eta$ is increasing with $q$, which implies that due to the dark energy effect, the virialized radius becomes larger than half of the turnaround radius. On the other hand, in the case where clustered dark energy cannot virialize, the virialized radius of the spherical over-dense region becomes smaller than half of the turn-around radius which is depicted by the green line in fig.~(\ref{etavsq1}). For both these cases, the equation of the state of dark energy is $\omega =-0.75$. Therefore, we can see that the presence of negative pressure in the dark energy fluid can create distinguishable large-scale structures of dark matter.

In \cite{Basilakos:2006us}, the authors considered a dynamical equation of state $\omega$ of clustered dark energy, where $\omega$ of clustered dark energy is a function of the scale factor $a$ of the over-dense regions. For this scenario, we can write down the following Friedmann equation for the background fluid:
\begin{eqnarray}
\frac{\ddot{\bar{a}}}{\bar{a}}=-\frac12\left[\left(\omega(\bar{a})+\frac13\right)\bar{\rho}_{DE}+\frac13\bar{\rho}_{DM}\right]\,\, ,
\label{F1}
\end{eqnarray}
where the background dark matter density $\bar{\rho}_{DM}\propto \bar{a}^{-3}$ and the background dark energy density $\bar{\rho}_{DE}\propto f(\bar{a})$, where
\begin{eqnarray}
f(\bar{a})=exp\left[3\int_{\bar{a}_0}^{\bar{a}}\left(\frac{1+\omega(\bar{a}_1)}{\bar{a}_1}\right)d\bar{a}_1\right]\,\, .
\end{eqnarray}
Inside the spherical over-dense regions, the Friedmann equation can be written as
\begin{eqnarray}
\frac{\ddot{a}}{a}=-\frac12\left[\left(\omega(a)+\frac13\right)\rho_{DE}+\frac13\rho_{DM}\right]\,\, ,
\label{F2}
\end{eqnarray}
where the $a$ is the scale factor of the over-dense region. In \cite{Basilakos:2006us}, the authors solved the Eqs.~(\ref{F1}, \ref{F2}), for different functional form of $\omega(a)$ and discussed the possible range of $\eta$. 

In all of the above scenarios, Friedmann equations predict that the homogeneous spherical over-dense regions collapse to a spacetime singularity at the time $t_c=2t_{max}$. However, before the singularity formation, the fluid inside the over-dense region reaches the virialization state and, in this way, cosmological large-scale structures are formed. The inclusion of the dark energy effects only shows how the negative pressure inside the dark energy changes the virialization radius and the density contrast of dark matter at the virialization state. Using general relativity, one can calculate the time when the over-dense region reaches the virialization state and also we can calculate the density contrast of dark matter at that virialization epoch. It is always true that a fluid system that has a non-zero total kinetic energy should always have a virialization state.  However, using general relativity, we cannot explain the virialization state. In order to explain the virialization state, we need to use the Newtonian theory of virialization. 

In the next section, we will use the general relativistic equilibrium condition of gravitational collapse to investigate the stable configurations in a cosmological scenario where dark energy can cluster inside the over-dense region of dark matter (i.e., one of the scenarios discussed above).  
\section{Equilibrium state with clustered dark energy}
\label{themodel}
\subsection{The Model}
In the standard top-hat collapse model, the fluid inside the spherical over-dense region is considered pressure-less (dust) and homogeneous \cite{GunnGott72}. Using this model, we can explain how spherical over-densities of cold dark matter evolves with time, where the dark energy effect is negligible. As we discussed before, there is literature where at the galaxy cluster scale, the effect of dark energy on the virialization of spherical over-dense regions of dark matter is investigated. In both homogeneous and clustered dark energy models, there exist two fluids (i.e., dark matter and dark energy) inside and outside of the spherical over-dense regions. Since the universe at a large scale is seemed to be spatially flat, the background spacetime seeded by the dark matter and dark energy outside the spherical over-dense region is generally considered as a spatially flat FLRW spacetime. On the other hand, the curvature of a dynamical spacetime inside the over-dense region depends upon the components of energy-momentum tensors of the two-fluid system. In our model, we consider clustered dark energy to avoid the energy conservation problems with the homogeneous dark energy discussed previously. Since we consider clustered dark energy, the spherical over-dense regions can be considered as isolated from the background universe. The dynamical spacetime seeded by the dark matter and the clustered dark energy inside the over-dense regions can have the following general form
\begin{eqnarray}
ds^2 = - e^{2\nu(r,t)} dt^2 + {R'^2\over G(r,t)}dr^2 + R^2(r,t) d\Omega^2\, ,
\label{genmetric2}
\end{eqnarray}  
where we have discussed the above dynamical spacetime in the previous section. Now, depending upon the components of the energy-momentum tensor, the above spacetime can be written in different forms, e.g., for a homogeneous fluid, the closed FLRW metric is one of the solutions of Einstein equations, and for an inhomogeneous, pressure-less fluid, the Lemaitre-Tolman-Bondi (LTB) spacetime is the well known dynamical solution of Einstein equations, etc. The energy-momentum tensor of the resultant fluid corresponding to the above metric can be written in terms of the energy-momentum tensors of component fluids (i.e., dark matter and clustered dark energy) in the following way
\begin{eqnarray}
T_{\mu \nu} &=& (T_{\mu \nu})_{DM}+(T_{\mu \nu})_{DE} \,\, \nonumber\\
&=&(P_{DM} + \rho_{DM})u_\mu u_\nu + P_{DM} g_{\mu \nu} \,\, \nonumber \\&+& (P_{DE}+
\rho_{DE})v_{\mu} v_\nu + P_{DE} g_{\mu \nu}\,,
\label{2femt}
\end{eqnarray}
where $\rho_{DM}$, $P_{DM}$, $u^{\mu}$ are the energy density, pressure, and four-velocity of dark matter, respectively, and $\rho_{DE}$, $P_{DE}$, $v^{\mu}$ are the energy density, pressure and four-velocity of clustered dark energy, respectively. As $u^{\mu}$ and $v^{\mu}$ are four velocities
$$u^\mu u_\mu = v^\mu v_\mu = -1\,.$$ If $u^{\mu}v_{\mu}\neq -1$, they are not comoving then the resultant fluid would have anisotropic pressures ($P_r \neq P_{\perp}$) \cite{Bhattacharya:2017chr, Dey:2019fja}. However, in our model, we consider the resultant fluid as a homogeneous and isotropic fluid. Therefore, in this paper we consider $u^{\mu}= v^{\mu}$. Hence, the energy density and pressure of the resultant fluid can be written as
\begin{eqnarray}
\rho &=& \rho_{DM}+\rho_{DE}\,\, ,\\
P &=& P_{DE}\,\, ,
\end{eqnarray}
where we consider the dark matter as pressure-less fluid (i.e., cold dark matter). Therefore, the equation of state of dark energy ($\omega_{DE}$) can be written as,
\begin{equation}
\omega_{DE}=\omega \left(\frac{1}{q}\left(\frac{a}{a_{max}}\right)^{-3}\frac{f(a)}{f(a_{max})}+1\right)\,\, ,
\label{wwde}
\end{equation} 
where $\omega$ is the equation of state of the resultant fluid and we consider $\rho_{DE}\propto f(a)$ inside the spherical over-dense region with the scale factor $a$. 

The dynamical spacetime, in Eq.~(\ref{genmetric2}), can reach the equilibrium state during its collapsing phase. In order to reach the equilibrium state, the equilibrium conditions stated in the previous section should be satisfied. 
In the next subsection, we show that the negative pressure and the time-varying equation of state of the dark energy help the total system to reach the equilibrium state. We construct a bonafide model of a homogeneous, non-dust, isotropic fluid collapse  that has an equilibrium end state. In order to show how the ratio between the scale factor at the equilibrium state ($a_e$) and the scale factor at the turnaround epoch ($a_{max}$) changes in our model, we consider the dark energy effect only in the collapsing phase. In our model, the expansion phase of the over-dense region is similar to the expansion phase of matter cloud in standard top-hat collapse model \cite{GunnGott72}. Therefore, we consider the total density $\rho \propto a^{-3}$ during the expansion phase and during the collapsing phase, $\rho \propto g(a)$, where
\begin{eqnarray}
g(a)=exp\left[3\int_{a_{max}}^{a}\left(\frac{1+\omega(a_1)}{a_1}\right)da_1\right]\,\, .
\label{ga1}
\end{eqnarray}
As the equation of state $\omega(a_{max})=0$, $g(a_{max})=a_{max}^{-3}$. Eq.~(\ref{ga1}) implies that the dark energy effect starts growing smoothly from the turnaround point. As we know, matter domination increases as we go from low redshift to high redshift. Therefore, we take a certain epoch ($a=a_{max}$) from where the effect of dark energy cannot be neglected. It should be noted that there are no particular reasons behind choosing that epoch ($a=a_{max}$). One can consider the dark energy effect throughout the evolution i.e., from the expansion phase to collapsing phase, and for that case, $g(a)=exp\left[3\int_{a_{0}}^{a}\left(\frac{1+\omega(a_1)}{a_1}\right)da_1\right]\,\,$.

In the first scenario, where the dark energy effect starts growing from the turnaround point, we have to match a matter-dominated closed FLRW metric with the dynamical metric seeded by an isotropic homogeneous fluid, at the spacelike hypersurface $t-t_{max}=0$. From the smooth matching of the metric in Eq.~(\ref{FLRWMetric}) and the metric in Eq.~(\ref{genmetric2}) at $t=t_{max}$, we get the following constraints \cite{Bhattacharya:2017chr, Dey:2019fja,Israel66}:
\begin{eqnarray}
&\nu(r,t_{\rm max})&=0,~f(r,t_{\rm max}) = a_{\rm max},~G(r,t_{\rm max}) = 1-r^2\, ,
\label{nmatch}\nonumber\\
& f(r, t_{\rm max})& \dot{G}(r,t_{\rm max}) = 2 \dot{f}(r,t_{\rm max})
 G(r,t_{\rm max}) = 0\,,
\label{cons1}
\end{eqnarray}   
where the last constrain implies that $\dot{f}(r,t_{\rm max})=\dot{G}(r,t_{\rm max})=0$.

In the next subsection, we present the bonafide homogeneous non-dust collapse model, where we consider the above matching conditions and the regularity conditions for the collapse discussed in the previous section.
\begin{figure*}
\includegraphics[width=185mm]{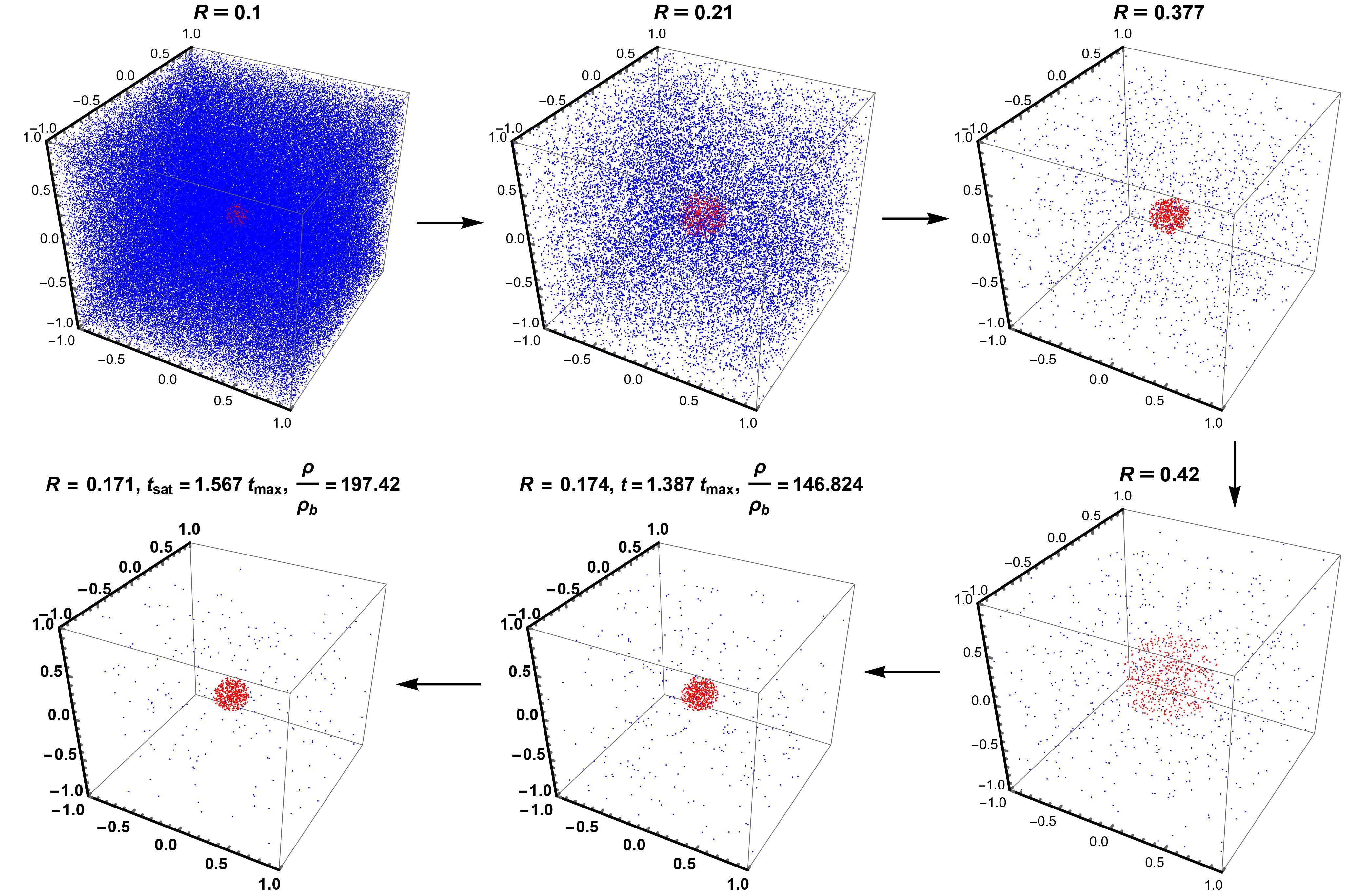}
 \caption{The figure depicts the evolution of the over-dense region in the dark matter field. The over-dense regions are shown by the spherically symmetric regions which are filled with red points and the background is shown by the region with blue points. Number of points per unit volume in both the regions describes the density of matter in those regions. It can be seen how the over-dense region expands first with the background (from the physical radius $R=0.1$ to $R=0.377$ ) and after the turn-around point (with physical radius $R=0.42$), how it collapses until the equilibrium state (with physical radius $R=0.171$) is achieved. The arrows in the figure indicate the direction of the evolution. The total number of points inside the spherical over-density is $10^3$, which is always conserved. This simulation is done, considering the bona-fide homogeneous non-dust collapse model.   }
 \label{bona1}
\end{figure*}
\subsection{An example of a bona-fide homogeneous Non-Dust collapse}
In this subsection, we present a simple model of gravitational collapse where, due to the simple form of the metric coefficients, we can analytically describe the dynamics of the collapsing matter from an initial state to the final equilibrium state.
From Eq.~(\ref{ms}), we can write
\begin{eqnarray}
    \dot{f}(r,t)=e^{\nu(r,t)}\left(\frac{F(r,t)}{r^3 f(r,t)}+\frac{G(r,t)-1}{r^2}\right)^{\frac12}.
    \label{ms1}
\end{eqnarray}
The solution of the above first-order, non-linear differential equation can give us information on the dynamics of the collapsing matter. The general integrated form of the solution of the above equation can be written as
\begin{eqnarray}
    t(r)=t_i + \int_f^1\frac{e^{-\nu(r,f)}}{\sqrt{\frac{F(r,f)}{r^3 f(r,f)}+\frac{G(r,f)-1}{r^2}}}df\,\, ,
    \label{intgen}
\end{eqnarray}
where the initial time is denoted by $t_i$. The above integrand is written as a function of $r$ and $f$, instead of $r$ and $t$ (since for a fixed comoving radius $r$, $f(r,t)$ monotonically decreases with time, one can write down all the components of the metric function as a function of $r$ and $f$). As stated before, one needs integrable forms of $\nu(r,f), G(r,f)$ and $F(r,f)$ to get an analytical result from the above integration. Therefore, in this subsection, using a simple form of the metric in Eq.~(\ref{genmetric}), we analytically show how a collapsing system reaches an equilibrium state satisfying all the conditions stated in Sec.~(\ref{nondust}).

In our simple collapsing model, we consider homogeneous matter with isotropic pressure. 
In order to construct this simple collapsing configuration, we consider  $G(r,t)$ to be time-independent during the collapsing phase (i.e., for $t>t_{max}$). Now, the matching conditions on $G(r,t)$ at $t=t_{max}$ (Eq.~(\ref{cons1})) fix the time-independent functional form of $G(r,t)$ $\forall t> t_{max}$:
\begin{eqnarray}
G(r,t)=1-r^2\,.
\label{gform}
\end{eqnarray}
Since
$G$ is considered to be time-independent, from Eq.~(\ref{vanond})
we readily get:
\begin{eqnarray}
\nu'(r,t) = 0\,\, ,
\end{eqnarray}
which implies $\nu(r,t)$ is a function of time only. Since $\nu(r,t)$ is now a function of time only, we can rescale the time coordinate in the general metric (Eq.~(\ref{genmetric})) and, therefore, one can always set 
\begin{eqnarray}
\nu(r,t) = 0\, .
\end{eqnarray}
The above expression of $\nu(r,t)$ satisfies the junction conditions stated in Eq.~(\ref{cons1}).
As mentioned above, during the collapsing phase we consider a homogeneous matter with isotropic pressure. To get the homogeneous configuration at the initial time of the collapse (i.e., at $t=t_{max}$), the functional form of Misner-Sharp mass has to be: $F(r,t_{max})=F_0(t_{max}) r^3$, where $F_0$ is a function of $t$ only. Since the expansion phase (i.e., for $t <t_{max}$) is modeled by the closed FLRW space-time, at the time $t=t_{max}$, the density of the matter is homogeneous and it can be written in terms of the maximum value of the scale factor ($a_{max}$) as:
\begin{eqnarray}
 \rho(t_{\rm max}) = {3\over a_{\rm max}^2}\,.
 \label{rhotm}
\end{eqnarray}
Therefore, from the expression of the energy density in Eq.~(\ref{eineqns}), we can readily get $F=a_{max}r^3$ which implies $F_0(t_{max})=a_{max}$. 
From the Eq.~(\ref{eineqns}), we can obtain the following expression
of pressure:
\begin{eqnarray}
 P =-\frac{\dot{F}}{\dot{R}R^2}={-r \dot f(1-G + r^2{\dot f}^2)- r f\dot G + 2 r^2 f\dot f
   \ddot R  \over r^3 f^2 \dot f}\,.
 \label{pres}
\end{eqnarray}
At the turning point $t=t_{max}$, the pressure becomes
\begin{eqnarray}
 P(t_{\rm max}) = {1\over a_{\rm max}^2} + 2 {\ddot f(t_{\rm max})\over a_{\rm max}}\,.
\end{eqnarray}
Since in our model, the matter is dust-like in the expansion phase, at the turning point the pressure should vanish. Hence, this gives the following constraint on $\ddot{f}$ at $t=t_{max}$:
\begin{eqnarray}
\ddot f(t_{\rm max}) = - \frac{1}{2a_{\rm max}}\, .
\label{p0cond} 
\end{eqnarray}
Therefore, in the contracting phase (i.e., when $t>t_{max}$), we get the following functional forms for $G(r,t), \nu(r,t)$ and $F(r,t)$: $G(r,t)=1-r^2, \nu(r,t)=0, F(r,t)=F_0(t)r^3$, which implies the contracting phase is also described by a closed FLRW spacetime. However, the main difference between the expanding and contracting phases is that in the contracting phase, the equation of state is non-zero while in the expanding phase, it is considered to be zero.
 As we stated before, in our model, the effect of clustered dark energy starts growing from the turnaround time ($t=t_{max}$). Therefore, during the collapsing phase, the resultant fluid acquires negative equation of state, where the relation between the equation of state ($\omega$) of the resultant fluid and the equation of state of dark energy ($\omega_{DE}$) is shown in Eq.~(\ref{wwde}). 

Up until now, we have considered the functional form of $G(r,t)$ which readily gives the radial independence of $\nu(r,t)$. Now, to solve the whole dynamics of the collapsing space-time, we need to choose the functional form of any one of the following functions: $F(r,t)$ or $R(r,t)$. As discussed above, we need to consider the Misner-Sharp mass in the following form: $F(r,t)=F_0(t)r^3$ to get a homogeneous configuration. Considering any suitable (i.e., it should follow the collapsing conditions stated above) form of $F_0(t)$, we can solve Eq.~(\ref{ms1}) to get the functional form of $f(r,t)$, and one can show that $f(r,t)$ becomes radially independent. For our present case, we need to consider $f(r,t)$ instead of $F(r,t)$, since we are interested in describing a collapsing scenario where homogeneous matter reaches an equilibrium state in an asymptotic comoving time. We know that at the turning point $t=t_{max}$, $\dot f=0, \ddot f<0 $ and at the equilibrium state $\dot f=0, \ddot f=0$.  Now, considering all these properties of the scale factor $f$, we propose the following possible form for $\dot f$:
\begin{eqnarray}
 \dot f(t_r)=-{A(t_r-1)\over(1+B(t_r-1)^2)^2}~,~t_r = {t\over t_{\rm max}}~,
\label{fdotexp}
\end{eqnarray}
\begin{figure*}
\includegraphics[width=185mm]{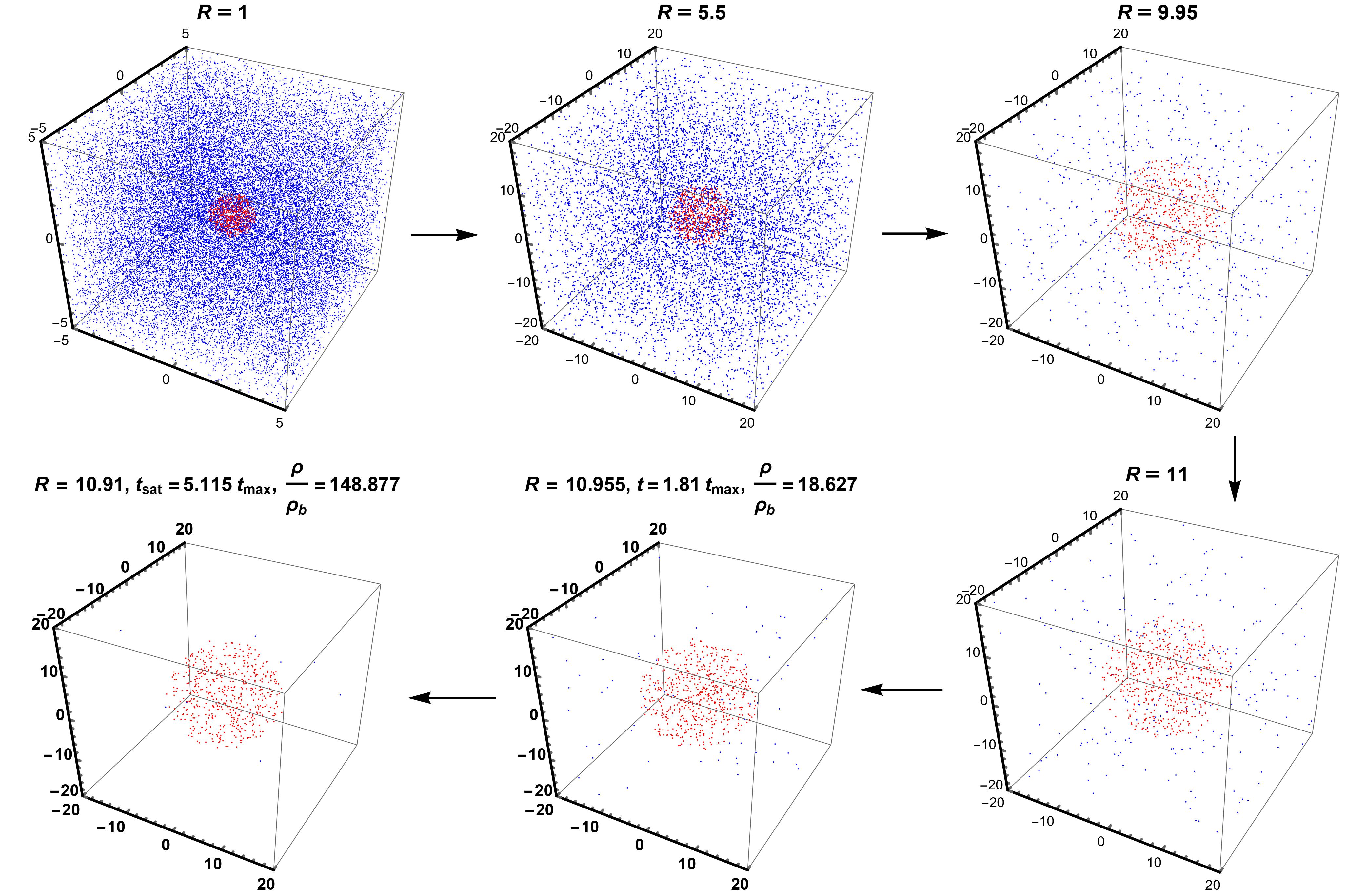}
 \caption{The figure shows the evolution of the over-dense region in the dark matter field, where the over-dense regions are shown by the spherically symmetric regions which are filled with red points and the background is shown by the region with blue points. The number of points per unit volume in both regions describes the density of matter in those regions. It can be seen how the over-dense region expands first with the background (from the scale factor $a=1$ to $a=9.95$ ) and after the turn-around point (with scale factor $a=11$), how it collapses until the equilibrium state (with scale factor $a=10.91$) is achieved. The arrows in the figure indicate the direction of evolution.  This simulation is done, considering the bona-fide homogeneous non-dust collapse model.   }
 \label{tophatevol2}
\end{figure*}
where it can be verified that the above expression of $\dot f$ satisfies the conditions at $t=t_{max}$ and at equilibrium state.
Here, $A$ and $B$ are positive real constants. Solving the above differential equation, we get the following form of the scale factor
$f$:
\begin{eqnarray}
 f(t_r)=f_e+\frac{A}{2 B \left(1+B (t_r-1)^2\right)}\, ,
\label{fex}
\end{eqnarray}
where $f_e$ is the value of $f$ at the equilibrium state. Now, we are in a position to write down the collapsing metric:
\begin{eqnarray}
ds^2 = - dt_r^2 + {f(t_r)^2\over (1-r^2)}dr^2 + r^2f(t_r)^2 d\Omega^2\, .
\label{eqbsptm}
\end{eqnarray}
Using the above metric, we get the following expressions of energy density and pressure:
\begin{widetext}
\begin{eqnarray}
\rho &=& \frac{3 \left(\frac{A^2 T^2}{\left(B T^2+1\right)^4}+1\right)}{\left(\frac{A}{2
   B \left(B T^2+1\right)}+f_e\right)^2}\,,\label{rhobonafide}\\
P &=& -\frac{4 B \left(A^2 \left(4 B T^2-1\right)+2 A Bf_e \left(3 B T^2-1\right) \left(B
   T^2+1\right)+B \left(B T^2+1\right)^4\right)}{\left(B T^2+1\right)^2
   \left(A+2 Bf_e \left(B T^2+1\right)\right)^2}\,,
\label{press}
\end{eqnarray}
\end{widetext}
where $T \equiv t_r - 1$. Since pressure is zero at the space-like hypersurface $t=t_{max}$ or $T=0$, we get
\begin{eqnarray}
f_e={B-A^2\over 2AB}\,,
\label{fzero}
\end{eqnarray}
which fixes the value of the scale factor at equilibrium. To get a positive definite value of the scale factor at equilibrium, we need $B>A^2$. 


\begin{figure*}
\centering
\begin{minipage}[b]{.4\textwidth}
\centering
\includegraphics[width=2.8in,height=2.5in]{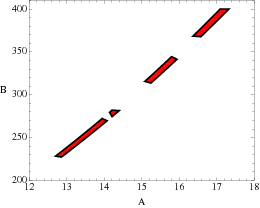}
\caption{The shaded regions depict the allowed values of parameters $A$ and $B$.}
\label{ABreg}
\end{minipage}\qquad
\begin{minipage}[b]{.4\textwidth}
\centering
\includegraphics[width=2.8in,height=2.8in]{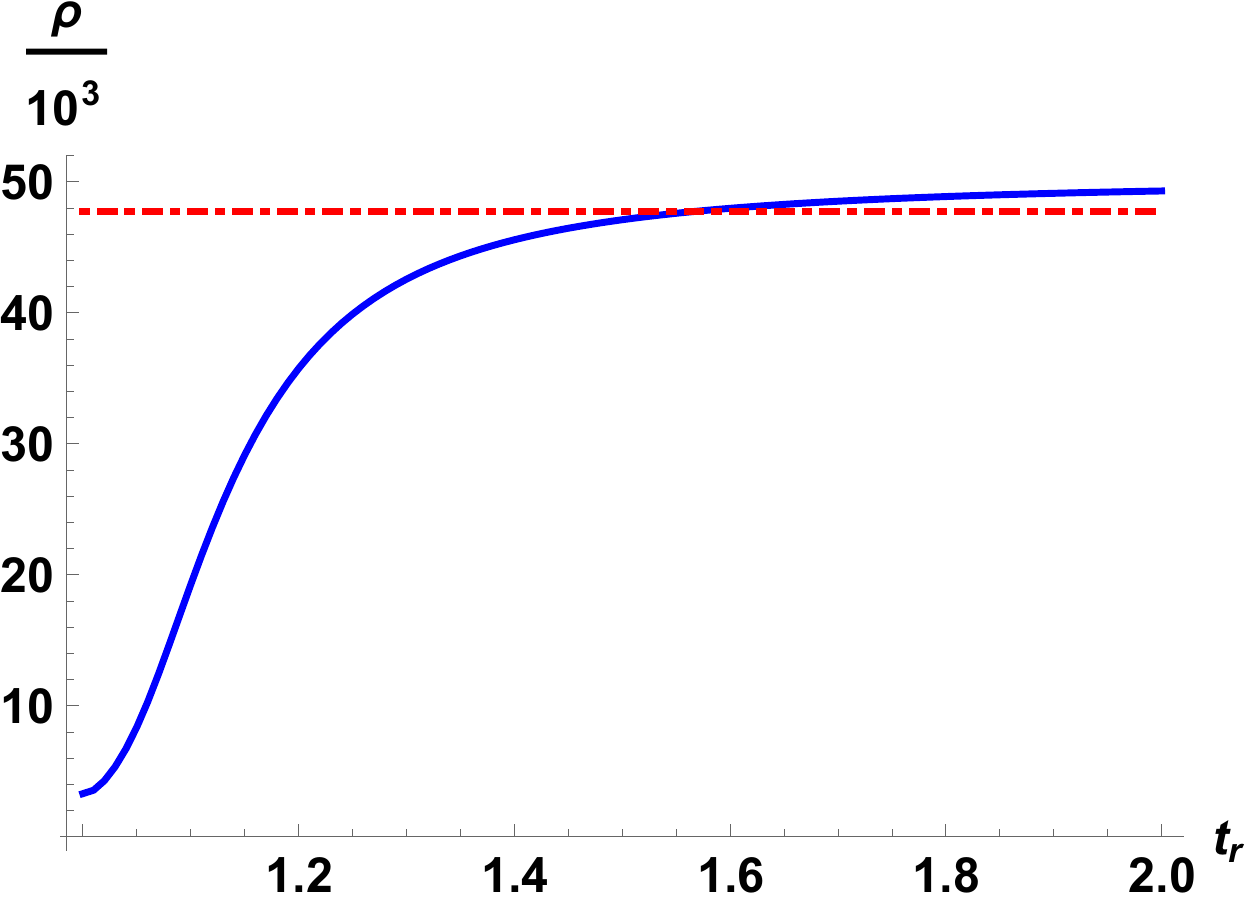}
\caption{Figure shows the nature of energy density ($\rho$) for $A =16.57$, $B= 369.3$. The dashed line indicates the density $\rho(t_{\rm sat})$ at the saturation time $t_{sat}$.}
\label{rhoAB}
\end{minipage}
\end{figure*}
Apart from the above-mentioned constraint on the $B$ and $A$, there might be other constraints on them which can come from all the collapse conditions discussed in section \ref{nondust}. We first verify whether the metric in Eq.~(\ref{eqbsptm}) satisfies the weak energy conditions. Using the expression of the scale factor at equilibrium in the expressions of energy density and pressure in Eqs.~(\ref{rhobonafide}),(\ref{press}), we get:
\begin{equation}
\rho = \frac{3 \left(\frac{A^2 T^2}{\left(B
   T^2+1\right)^4}+1\right)}{\left(\frac{B-A^2}{2 A B}+\frac{A}{2 B \left(B
   T^2+1\right)}\right)^2}~,~
\label{rhof}
\end{equation}
and
\begin{widetext}
\begin{eqnarray}
\rho + P = \frac{4 A^2 \left(A^2 T^2 \left(3 B T^2+1\right)+\left(B T^2+1\right) \left(B
   T^2 \left(2 B T^2 \left(B T^2+3\right)+3\right)+3\right)\right)}{\left(B
   T^2+1\right)^2 \left(-A^2 T^2+B T^2+1\right)^2}\,.
\label{WEC1}
\end{eqnarray}
\end{widetext}
The above expressions of $\rho$ and $\rho +P$ show that they are always positive for all positive values of $A, B$, which implies that during the collapsing phase the weak energy conditions (Eq.(\ref{WEC1})) are always satisfied.
During the collapsing phase, we consider a homogeneous configuration that manifestly satisfies the regularity conditions. Due to the homogeneity, $\rho$ and $P$ are regular at $r=0$.
Finally, as discussed above, to avoid the shell-crossing scenario we require $R'(r,t)=f(r,t)+r f'(r,t)> 0$ which ensures the absence of shell-crossing singularity. Now, since in the above bona-fide model of collapse the scale factor $f$ is independent of the radial coordinate, $R'(r,t)>0 \implies f(t)>0$. 
Therefore, using $T = t_r -1$, we can write 
\begin{equation}
f = \frac{1}{2 A}-\frac{A T^2}{2 \left(B T^2+1\right)} > 0\,.
\label{regcon1}
\end{equation}
It can be verified that for $f$ to be remain positive as $t_r\to \infty (\equiv
T\to \infty)$, we need $B > A^2$, which is the condition discussed previously.

From the expression of $\dot f$ in Eq.~(\ref{fdotexp}), it can be verified that $\dot f$ is zero at $t=t_{max}$ and it also tends to zero in large comoving time ($\mathcal{T}$) and in between (i.e., $t_{max}\leq t<\mathcal{T}$) it has negative values. Therefore, it is obvious that there exists a turning point of $\dot f$ in the time interval: $t_{max}\leq t<\mathcal{T}$, where $\ddot f =0$. After differentiating $\dot f$ (Eq.~(\ref{fdotexp})) with respect to $t$, we can write 
\begin{equation}
{\ddot f(t_r)} = 0 \Rightarrow T = \frac{1}{\sqrt{3B}}~.
\label{tpfdot}
\end{equation}
Therefore, in the interval $\frac{1}{\sqrt{3B}}\leq T < \infty$, the absolute value of $\dot f$ decreases gradually and becomes zero in asymptotic comoving time. One can verify that the violation of strong energy condition (SEC) causes the deceleration of the collapsing velocity of matter from the time $T= \frac{1}{\sqrt{3B}}$. In order to satisfy strong energy condition, we need $R_{\mu\nu}u^{\mu}u^{\nu}\geq 0$, where $u^{\mu}$ is the temporal orthonormal basis of a frame \cite{Poisson} and that implies: $\rho+ 3P \geq 0$ which translates into:
\begin{equation}
\frac{12A^2B\left(1 - 3BT^2\right)}{\left(1 + BT^2\right)^2\left(A^2 + \left(B - A^2\right)\left(1 + BT^2\right)\right)} >0 ~.
\label{SEC}
\end{equation}
Since $B>A^2$, it can be verified that the above inequality does not hold in the time interval $\frac{1}{\sqrt{3B}}\leq T < \infty$ where $\ddot f>0$. Therefore, violation of SEC slows down the collapsing dynamics and the collapsing system reaches an equilibrium configuration in an asymptotic comoving time. As mentioned previously, in our model, the resultant fluid is a combination of two fluids: dark matter and clustered dark energy. Therefore, due to the dark energy, the total fluid violates the strong energy condition.
\begin{figure*}
\centering
\subfigure[Evolution of the equation of state $\omega$ of the resultant fluid with the scale factor $a$, during collapsing phase, where $A=16.57, B=369.3$ and $\Omega_{m0}=1.31556$.]
{\includegraphics[width=80mm]{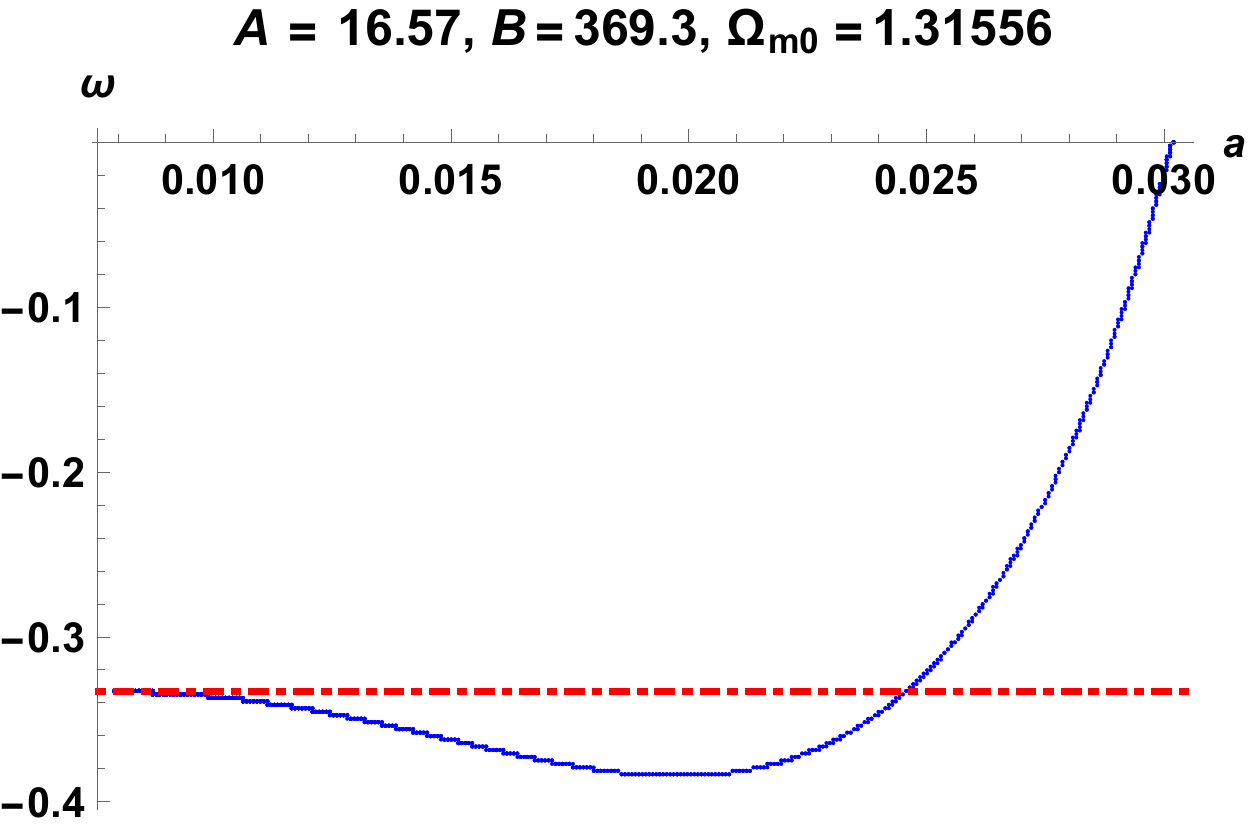}\label{eosbona1}}
\hspace{0.1cm}
\subfigure[Evolution of the equation of state $\omega$ of the resultant fluid with the scale factor $a$, during collapsing phase, where $A=0.122, B=1.143$ and $\Omega_{m0}=1.1$.]
{\includegraphics[width=80mm]{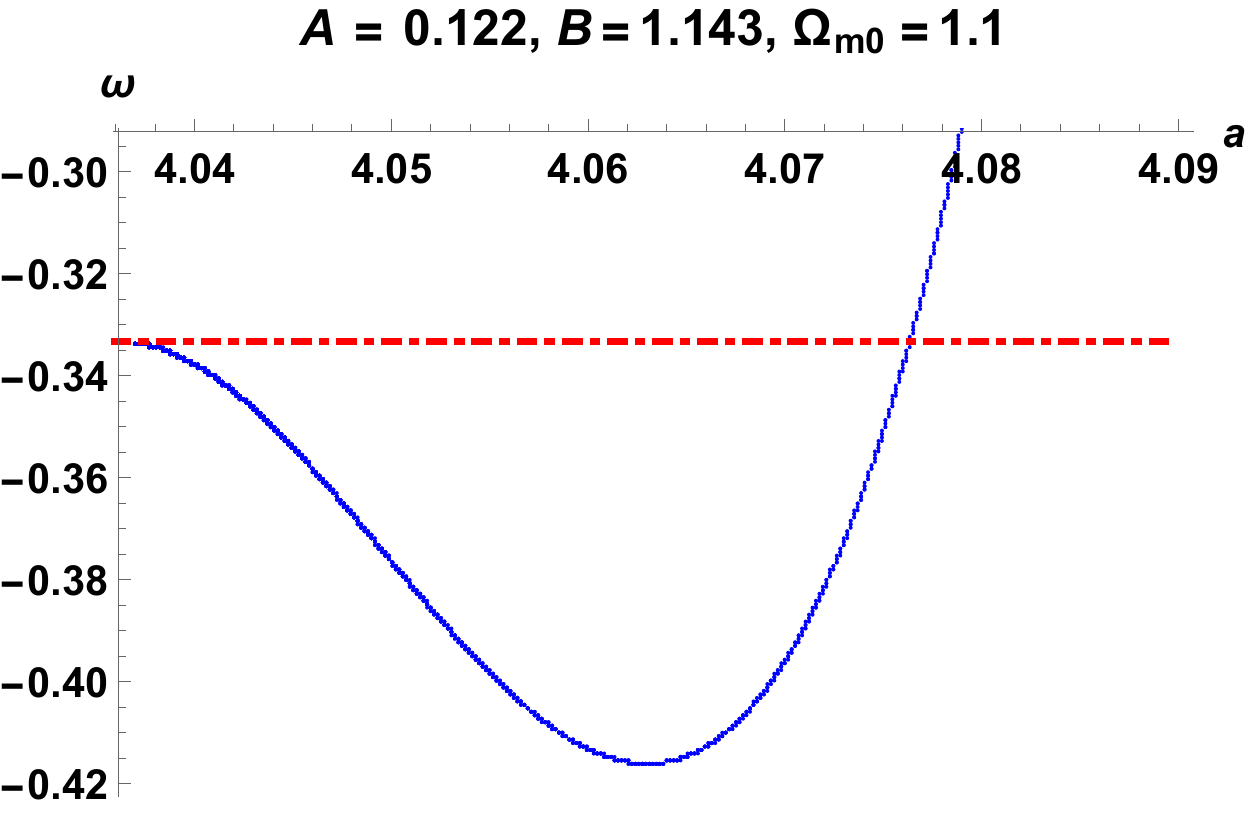}\label{eosbona2}}
 \caption{Figure shows how the equation of state ($\omega$) varies with the scale factor $a$ during the collapsing phase of spherical over-dense regions.  During the collapse, scale factor ($a$) decreases with time, therefore, in both above figures $\omega$ evolves from $0$ (at $t=t_{max}$) to $-\frac13$ (at $t=t_{sat}$). }
 \label{wvsq}
\end{figure*}
\subsection{ Comparison with spherical top-hat collapse}
In the previous subsection, we have discussed how the scale factor of a collapsing homogeneous over-dense region reaches its equilibrium value $f_e = \frac{B-A^2}{2AB}$, where the initial value (i.e., at $t=t_{max}$) of the scale factor is: $f_{max}=\frac{1}{2A}$. Since the collapsing system reaches the equilibrium configuration in asymptotic time, we develop the following algorithm to compare our model with the standard top-hat collapse model.
At the time $t= t_{ max}$
and $t\to \cal{T}$, the energy densities of the over-dense matter distribution become
\begin{eqnarray}\label{rhoe}
\rho(t_{\rm max}) = 12A^2,~~ \rho_e= \frac{12 A^2}{\left(1-\frac{A^2}{B}\right)^2}\,\, ,
\label{rhomaxe}
\end{eqnarray}
where $\rho_e$ is the equilibrium density. The equilibrium density is $\frac{1}{\left(1-\frac{A^2}{B}\right)^2}$ times greater than the density at turnaround time.  Now, we introduce a
time, $t_{\rm sat}$, which we call saturation time. At that time, the difference
between the turnaround density $\rho(t_{\rm
  max})$ and $\rho(t_{\rm sat})$ reaches $95\%$ of the difference
between $\rho(t_{\rm max})$ and $\rho_e$. Therefore, at $t=t_{sat}$
\begin{widetext}
\begin{eqnarray}
  \rho(t_{\rm sat}) = \rho(t_{\rm max}) + (\rho_e -\rho(t_{\rm max}))\times 0.95 =
  \frac{3 A^2 \left(A^4-2 A^2 B+20 B^2\right)}{5 \left(A^2-B\right)^2}~ .
\label{densat}
\end{eqnarray}
\end{widetext}
This definition of the saturation time mentioned above allows us to predict the various properties of the collapsing system at the finite saturation time. Since the density of the collapsing system reaches $95\%$ of the equilibrium density, one can represent the equilibrium state by the saturation state with the $95\%$ or $2\sigma$ accuracy.
As discussed in the previous section, in the spherical top-hat collapse model, the linear density contrast at the virialization state is
\begin{eqnarray}
  \delta_{\rm vir} = \left(\frac{\Delta\rho}{\bar{\rho}}\right)_{\rm critical}=
        {3\over 20} \left(6\pi {t_{\rm vir}\over t_{\rm max}}\right)^{2/3} \sim 1.686\,.
\nonumber        
\end{eqnarray}
Now, we assume the $t_{sat}$ in our model is analogous to the $t_{\rm vir}$ in top-hat collapse model. With this assumption, the linear density contrast at the saturation time $t_{sat}$ can be written as:
\begin{eqnarray}
\delta_{\rm sat} ={3\over 20} \left(6\pi {t_{\rm sat}\over t_{\rm max}}\right)^{2/3}\,.
\label{critdentr}
\end{eqnarray}
The reason behind the similarity in the expressions of the linear density contrast in our model and the top-hat collapse model is that the both models have the same origin (i.e.,
the time when the over-dense region starts expanding) and the expansion phase of the both models are exactly similar. As we know, in our model, pressure becomes non-zero during the collapsing phase which changes the collapsing dynamics. The different results obtained from these two similar formulas are related to the two different timescales; i.e., in our model
$t_{sat}$ which may or may not be equal to $t_{vir}$. 

As we know, in the top-hat collapse model, the linear density contrast at virialization $\delta_{vir}\sim 1.686$ and the corresponding density ratio $  \frac{\rho(t_{\rm vir})}{\bar{\rho}(t_{\rm vir})} \sim
  170 - 200\,$. Therefore, in order to compare our model with the top-hat model, we investigate whether at the saturation time $t_{sat}$ the collapsing over-dense region in our model attains similar values of density contrast and ratio. Considering at $t=t_{sat}$
\begin{eqnarray}
\delta_{sat}&\sim& 1.686,\\
  \frac{\rho(t_{\rm sat})}{\bar{\rho}(t_{\rm sat})} &\sim&
  170 - 200\,,
\label{ratio4}
\end{eqnarray}  
we would find the parameters' space of $A, B$ for which the above-mentioned conditions are satisfied.

Before proceeding further, we would like to write down the density ratio given in
Eq.~(\ref{ratio4}) as a function of $t_{\rm sat}/ t_{\rm max}$: 
\begin{widetext}
\begin{eqnarray}
{\rho(t_{\rm sat})\over\bar{\rho}(t_{\rm sat})}& =&{\rho(t_{\rm
    sat})\over\rho(t_{\rm max})}\times 
{\rho(t_{\rm max})\over \bar{\rho}(t_{\rm max})}\times
{\bar{\rho}(t_{\rm max})\over \bar{\rho}(t_{\rm sat})} 
 \nonumber\\
 &= & 5.55\times{\rho(t_{\rm sat})\over\rho(t_{\rm max})}
 \frac{\bar{a}^3(t_{\rm sat})}{\bar{a}^3(t_{\rm max})}=
5.55\times{\rho(t_{\rm sat})\over\rho(t_{\rm max})}\left(\frac{t_{\rm
    sat}}{t_{\rm max}}\right)^2\,,
\label{rhocon1}    
\end{eqnarray}
\end{widetext}
where we use the fact that the density of the over-dense region is 5.55 times greater than the density of the background at
$t=t_{max}$. We assume the background energy density is proportional to $\bar{a}(t)^{-3}$, where $\bar{a}(t)\propto t^{2/3}$. The reason behind this assumption is that we want to obtain the minimum possible value of density ratio $\rho(t_{\rm sat})\over\bar{\rho}(t_{\rm sat})$. One can consider the dark energy effect in the background which would give us a larger value of that density ratio at the saturation time. If we consider the dark energy effect in the background expansion then the functional form of the background scale factor $\bar{a}$ can be obtained from Eq.~(\ref{F1}).
The above expression of the density ratio at $t= t_{sat}$ as a function of $t_{sat}/ t_{max}$ will be useful in our analysis. 

Now, in the above equation, $\rho(t_{max})$ is known from Eq.~(\ref{rhomaxe}) in terms of $A$ and $\rho(t_{sat})$ can be calculated from the Eq.~(\ref{densat}). As we discussed before, in order to compare with the top-hat collapse model, we need to constrain the density ratio $\frac{\rho(t_{sat})}{\bar{\rho}(t_{sat})}$ in the range $170-200$ and the linear density contrast $\delta_{sat}\sim 1.69$.  Therefore, we need to find out the parameters' space of $A$ and $B$ for which $\delta_{sat}\sim 1.69$, $170 <\rho(t_{\rm
  sat})/\bar{\rho}(t_{\rm sat})<205$, $1<t_{\rm sat}/t_{\rm max}<2$ and along with the condition $B >
A^2$ (from Eq.(\ref{fzero}). As we know, the last condition is necessary to make the energy density of the collapsing matter positive.  
In fig.~(\ref{ABreg}), we depict the allowed regions of the parameters $A$
and $B$, where the above constraints are satisfied. In fig.~ (\ref{rhoAB}), we show how the
density $\rho$ of Eq.(\ref{rhof}) changes with the proper time, where the dotted red line indicates the value of
$\rho(t_{sat})$ defined in Eq.(\ref{densat}). For better visibility, in fig.~ (\ref{rhoAB}), we scale $\rho$ and $\rho(t_{\rm sat})$ by a factor of $10^3$. In fig.~ (\ref{rhoAB}), we consider $A=16.57$ and $B=369.3$ which is taken from the allowed region of $A$ and $B$ shown in fig.~(\ref{ABreg}). 

In fig.~(\ref{bona1}), with the parameter values $A=16.57, B=369.3$ and $\Omega_{m0}=1.311556$,  we show how the spherical over-dense region in dark matter field evolves with time, where the dark energy effect during collapsing phase is considered. One can verify that the expansion phase in fig.~(\ref{bona1}) is similar to the expansion phase shown in fig.~(\ref{tophatevol}). However, during the collapsing phase, they differ from each other. In fig.~(\ref{tophatevol}), it can be seen that the virialization radius ($R_{vir}$) is half of the turnaround radius ($R_{max}$). On the other hand, in fig.~(\ref{bona1}), one can verify that the ratio between the equilibrium radius and the turnaround radius is less than half. Therefore, we can say that due to the dark energy effect, the spherical over-dense region has to contract more to reach the static state. In subsection (\ref{virwithdark}), we discuss the dark energy effect in the virialization of the dark matter, where the behavior of $\eta=\frac{R_{vir}}{R_{max}}$ depicted in fig.(\ref{etavsq1}) show similar type of results what we get for the bona-fide model. During the collapsing phase, the equation of state of the resultant fluid ($\omega$) varies with scale factor $a$, and at the saturation time ($t_{sat}$) it becomes $-\frac13$. The behaviour of $\omega$ is depicted in fig.~(\ref{eosbona1}), where $A=16.57, B=369.3$ and $\Omega_{m0}=1.311556$.

For $A=0.122, B=1.143$ and $\Omega_{m0}=1.1$, the density ratio between the density of the over-dense region and the density of the background becomes $148.877$ and the ratio between equilibrium radius and the turnaround radius becomes greater than half which is shown in fig.~(\ref{tophatevol2}). In this case, the behavior of the equation of state of the resultant fluid ($\omega$) is shown in fig.~(\ref{eosbona2}), where it can be seen that like the previous case, the equation of state becomes $-\frac13$ at the saturation time.

\section{conclusion}
\label{conclude}
The concluding remarks on the results of the present paper are as follows:
\begin{itemize}
    \item In section (\ref{nondust}), after discussing the previous work by Joshi et al. \cite{JMN11, Joshi:2013dva} on the equilibrium conditions of gravitational collapse of spherically symmetric matter, we first show how the relativistic covariant form the Eq.~(\ref{inertia1}) implies the equilibrium conditions of the gravitational collapse of a homogeneous configuration with a constant equation of state, i.e., $
\lim_{t\to\mathcal{T}}\dot{R}=\lim_{t\to\mathcal{T}}\ddot{R}=0\,\, ,
$
where the limit is taken for a large comoving time  $\mathcal{T}$. As we know, Eq.~(\ref{inertia1}) is the Newtonian virialization condition of a collapsing self-gravitating system of particles. However, as mentioned before, we don't know whether its relativistic covariant form (i.e., $\lim_{\tau \to \cal{T}}u^{\alpha}\nabla_{\alpha}u^{\beta}\nabla_{\beta}I=0$ ) is the general relativistic condition of virialization. Further investigation is necessary in this direction. 
\item In section (\ref{nondust}), we discuss the spinor structure of the general collapsing space-time (Eq.~(\ref{genmetric})). Using Cartan-Karlhede algorithm, we redefine the equilibrium condition using the Cartan scalars: $D\Psi_2$ and $\Delta \Psi_2$. We show that $(D-\Delta)^2\Psi_2=0~ \forall r$ implies $\dot{R}=\ddot{R}=0\,\, $. Therefore, we can define the equilibrium condition as  $$\lim_{t \to \cal{T}}(D-\Delta)^2\Psi_2 = 0~ \forall r.$$
\item In section (\ref{tophat}), we discuss the standard model of top-hat collapse and review the works where the top-hat model is modified considering the dark energy effect inside the over-dense region of dark matter. In section (\ref{themodel}), we use the equilibrium condition of gravitational collapse (Eq.~(\ref{eqlb})) in a similar cosmological scenario mentioned above where dark energy can cluster inside the over-dense region of dark matter. In order to construct a pedagogical model of a system consisting of dark matter and dark energy, we construct a two-fluid system where one fluid is pressureless and the other one has negative pressure. Then, we propose a collapsing space-time (Eq.~(\ref{eqbsptm})) that can be seeded by the two-fluid system and show that the spacetime has an equilibrium configuration at an asymptotic comoving time. In our model, we consider a non-zero dark energy effect after the turnaround time ($t_{max}$) and, therefore, before the turnaround time; the dynamics of the over-dense region in our model are similar to that in the top-hat collapse model.
\item For a more general scenario of gravitational collapse, where the collapsing matter is spatially inhomogeneous, one can investigate the possible equilibrium state that can be achieved by the collapsing system in an asymptotic co-moving time. In order to get a full analytic solution of the dynamics of such a type of gravitational collapse, one needs to have an integrable functional form of the integrand in Eq.~(\ref{intgen}). As mentioned before, in this paper we consider the simplest case where the collapsing matter is homogeneous. An inhomogeneous collapsing scenario having an end equilibrium state will be discussed elsewhere.
\item In section (\ref{themodel}), we compare the properties of the equilibrated configuration with that of a virialized system. We first define a saturation time ($t_{sat}$); i.e., the time when the difference
between the turnaround density $\rho(t_{\rm
  max})$ and $\rho(t_{\rm sat})$ reaches $95\%$ of the difference
between $\rho(t_{\rm max})$ and $\rho_e$, where $\rho_e$ is density at equilibrium. Since a collapsing system can reach the equilibrium configuration in an asymptotic comoving time whereas it can reach the virialized state in a finite time ($t_{vir}$), the introduction of $t_{sat}$ is very important in order to compare the two different stable configurations. From fig.~(\ref{rhoAB}), it can be seen that initially the density  of collapsing system increases very rapidly compared to the increment of density after $t_{sat}$. Therefore, a collapsing system reaches the equilibrium configuration with a $2\sigma$ level accuracy (i.e., the $95\%$ accuracy). After defining $t_{sat}$, we compare the equilibrium radius $r_{sat}$ with the virialized radius $r_{vir}$ of a collapsing system. As we know, in the top-hat collapse model, a collapsing system virializes at time $t_{vir}=1.81 t_{max}$ and at that time the virialized radius is half of the turnaround radius. From fig.~(\ref{bona1}), it can be seen that for some parameters' values, at the time $t_{sat}= 1.567 t_{max}$ the radius of the equilibrated system is less than half of the turnaround radius. On the other hand, it can also be shown that there is a parameter space for which the equilibrated radius is greater than half of the turnaround radius (fig.~(\ref{tophatevol2})). For the above-mentioned two cases, the following inequality always holds $170 <\rho(t_{\rm
  sat})/\bar{\rho}(t_{\rm sat})<205$, since this interval of density ratio has cosmological relevance.
\end{itemize}
As discussed before, a spherically symmetric dust collapse cannot reach the equilibrium state. Therefore, the general relativistic equilibrium conditions may play an important role in the structure formation at a certain cosmological scale where the negative pressure of dark energy influences the collapsing dynamics of the over-dense regions of dark matter. This paper only addresses this possibility. A detailed investigation of the possible observational signatures of the equilibrium configuration at a certain cosmological scale is left for future work.

\section{Acknowledgement}

DD would like to acknowledge the support of
the Atlantic Association for Research in the Mathematical Sciences (AARMS) for funding the work. DD also would like to acknowledge Dr. Kaushik Bhattachariya, Dr. Tapobrata Sarkar, and Dr. Arindam Mazumdar for the fruitful discussion on the work. AC acknowledges financial support from NSERC.


\begin{thebibliography}{99}
\bibitem{Lynden-Bell67}
D. Lynden-Bell,
\href{http://adsabs.harvard.edu/full/1967MNRAS.136..101L}{Mon. Not. Roy. Astron. Soc. {\bf 136}, 101 (1967).}

\bibitem{Merritt99} 
D. Merritt,
\href{http://iopscience.iop.org/article/10.1086/316307/pdf}{Publ. Astron. Soc. Pac. {\bf 111}, 129 (1999).}


\bibitem{GunnGott72}
J. E. Gunn and J. R. Gott III, 
\href{http://adsabs.harvard.edu/doi/10.1086/151605}{ApJ {\bf 176}, 1 (1972).}

\bibitem{virialGR1}
S. Bonazzola,
\href{https://adsabs.harvard.edu/full/1973ApJ...182..335B}{Astrophysical Journal, {\bf 182}, 335-340 (1973).}

\bibitem{virialGR2}
C. Vilain
\href{https://adsabs.harvard.edu/full/1979ApJ...227..307V}{Astrophysical Journal, {\bf 227}, 307-318 (1979).}

\bibitem{virialGR3}
E. Gourgoulhon and S. Bonazzola
\href{https://iopscience.iop.org/article/10.1088/0264-9381/11/2/015/meta}{Class. Quantum Grav. {\bf 11}, 443 (1994).}

\bibitem{virialGR4}
S.~Meyer, F.~Pace and M.~Bartelmann,
\href{https://journals.aps.org/prd/abstract/10.1103/PhysRevD.86.103002}{Phys. Rev. D \textbf{86}, 103002 (2012).}

\bibitem{JMN11} 
P. S. Joshi, D. Malafarina, and R. Narayan, 
\href{http://iopscience.iop.org/article/10.1088/0264-9381/28/23/235018/meta}{Class. Quantum Grav. {\bf 28}, 235018 (2011).}

\bibitem{Joshi:2013dva}
P.~S.~Joshi, D.~Malafarina and R.~Narayan,
\href{https://iopscience.iop.org/article/10.1088/0264-9381/31/1/015002}{Class. Quant. Grav. \textbf{31}, 015002 (2014).}


\bibitem{Bhattacharya:2017chr}
K.~Bhattacharya, D.~Dey, A.~Mazumdar and T.~Sarkar,
\href{https://journals.aps.org/prd/abstract/10.1103/PhysRevD.101.043005}{Phys. Rev. D \textbf{101}, no.4, 043005 (2020).}

\bibitem{Dey:2019fja}
D.~Dey, P.~Kocherlakota and P.~S.~Joshi,
\href{https://arxiv.org/abs/1907.12792}{arXiv:1907.12792 [gr-qc].}

\bibitem{MisnerSharp64} 
C. W. Misner and D. H. Sharp,
\href{https://journals.aps.org/pr/abstract/10.1103/PhysRev.136.B571}{Phys. Rev., \textbf{136}, 2B (1964).}


\bibitem{Lahav:1991wc}
O.~Lahav, P.~B.~Lilje, J.~R.~Primack and M.~J.~Rees,
\href{https://ui.adsabs.harvard.edu/abs/1991MNRAS.251..128L}{Mon. Not. Roy. Astron. Soc. \textbf{251}, 128-136 (1991).}

\bibitem{Steinh}
L.~Wang, P.~J.~Steinhardt,
\href{https://iopscience.iop.org/article/10.1086/306436}{
IOP Publishing, \textbf{508},2, 483-490 (1998).}

\bibitem{Shapiro}
I.~T.~Iliev, P.~R.~Shapiro,
\href{https://academic.oup.com/mnras/article/325/2/468/1155673}{MNRS, \textbf{325},2, 468-482 (2001).}

\bibitem{Basilakos:2003bi}
S.~Basilakos,
\href{https://ui.adsabs.harvard.edu/abs/2003ApJ...590..636B}{Astrophys. J. \textbf{590}, 636-640 (2003).}



\bibitem{Caldwell:2003vq}
R.~R.~Caldwell, M.~Kamionkowski and N.~N.~Weinberg,
\href{https://journals.aps.org/prl/abstract/10.1103/PhysRevLett.91.071301}{Phys. Rev. Lett. \textbf{91}, 071301 (2003).}

\bibitem{Horellou:2005qc}
C.~Horellou and J.~Berge,
\href{https://academic.oup.com/mnras/article/360/4/1393/1079075}{Mon. Not. Roy. Astron. Soc. \textbf{360}, 1393-1400 (2005).}

\bibitem{Maor:2005hq}
I.~Maor and O.~Lahav,
\href{https://iopscience.iop.org/article/10.1088/1475-7516/2005/07/003}{JCAP \textbf{07}, 003 (2005).}

\bibitem{Percival:2005vm}
W.~J.~Percival,
\href{https://www.aanda.org/articles/aa/abs/2005/45/aa3637-05/aa3637-05.html}{Astron. Astrophys. \textbf{443}, 819 (2005).}

\bibitem{mota2006}
N.~J.~Nunes, D.~F.~Mota,
\href{https://academic.oup.com/mnras/article/368/2/751/985027}{MNRS, \textbf{368},2, 751-758 (2006).}

\bibitem{Wang:2005ad}
P.~Wang,
\href{https://iopscience.iop.org/article/10.1086/500074}{Astrophys. J. \textbf{640}, 18-21 (2006).}

\bibitem{Basilakos:2006us}
S.~Basilakos and N.~Voglis,
\href{https://academic.oup.com/mnras/article/374/1/269/961317}{Mon. Not. Roy. Astron. Soc. \textbf{374}, 269-275 (2007).}

\bibitem{Maor:2006rh}
I.~Maor,
\href{https://link.springer.com/article/10.1007/s10773-007-9344-z}{Int. J. Theor. Phys. \textbf{46}, 2274-2282 (2007).}

\bibitem{Basilakos:2009mz}
S.~Basilakos, J.~Bueno Sanchez and L.~Perivolaropoulos,
\href{https://journals.aps.org/prd/abstract/10.1103/PhysRevD.80.043530}{Phys. Rev. D \textbf{80}, 043530 (2009).}

\bibitem{Basilakos:2010rs}
S.~Basilakos, M.~Plionis and J.~Sola,
\href{https://journals.aps.org/prd/abstract/10.1103/PhysRevD.82.083512}{Phys. Rev. D \textbf{82}, 083512 (2010).}

\bibitem{Leewang}
S.~Lee, Kin-Wang Ng,
\href{https://iopscience.iop.org/article/10.1088/1475-7516/2010/10/028}{JCAP, \textbf{2010},10, 028-028, (2010).}

\bibitem{Chang:2017vhs}
C.~C.~Chang, W.~Lee and K.~W.~Ng,
\href{https://linkinghub.elsevier.com/retrieve/pii/S2212686417300675}{Phys. Dark Univ. \textbf{19}, 12-20 (2018).}

\bibitem{OppenheimerSnyder39} 
J. R. Oppenheimer and H. Snyder,
\href{https://journals.aps.org/pr/abstract/10.1103/PhysRev.56.455}{Phys. Rev. {\bf 56}, 455 (1939).}

\bibitem{Datt} 
S. Datt, 
Zs. f. Phys. {\bf 108}, 314 (1938).


\bibitem{Israel66}
W. Israel, 
\href{https://link.springer.com/article/10.1007/BF02710419}{Nuovo Cimento B {\bf 44}, 1 (1966).}

\bibitem{Meyer:2012nw} 
  S.~Meyer, F.~Pace and M.~Bartelmann,
\href{https://journals.aps.org/prd/abstract/10.1103/PhysRevD.86.103002}{  Phys.\ Rev.\ D {\bf 86}, 103002 (2012).}  

\bibitem{Creminelli:2009mu} 
  P.~Creminelli, G.~D'Amico, J.~Norena, L.~Senatore and F.~Vernizzi,
\href{https://iopscience.iop.org/article/10.1088/1475-7516/2010/03/027/meta}{JCAP {\bf 1003}, 027 (2010).}

\bibitem{Hellaby}
 C.~ Hellaby and K.~ Lake,
 \href{http://adsabs.harvard.edu/full/1985ApJ...290..381H}{Astrophys.\ J. \textbf{290} 381 (1985).}
 
\bibitem{Szekeres}
P.~ Szekeres and A. Lun,
\href{https://core.ac.uk/download/pdf/191914758.pdf}{J. Austral. Math. Soc.B \textbf{41}, 167–179 (1999). }

\bibitem{Joshi:2012ak}
P.~S.~Joshi and R.~V.~Saraykar,
\href{https://www.worldscientific.com/doi/abs/10.1142/S0218271813500272}{Int. J. Mod. Phys. D \textbf{22}, 1350027 (2013).}

\bibitem{Poisson}
E.~ Poisson,
\href{https://www.cambridge.org/core/books/relativists-toolkit/DA7ED68B971708A0F782257F948981E7}{CambridgeUniversity Press, 2004.}

\bibitem{Adams1}
R.C.~Adams, R.D.H~Warburton and J.M.~Cohen,
\href{https://ui.adsabs.harvard.edu/link_gateway/1975ApJ...200..263A/doi:10.1086/153784}{Astrophysical Journal \textbf{200}, 263-268 (1975).}
\bibitem{Adams2}
R.C.~Adams, J.M.~Cohen, R.J. Adler, and C.~Sheffield,
\href{https://doi.org/10.1103/PhysRevD.8.1651}{Phys. Rev. D \textbf{8}, 1651 (1973).}
\bibitem{Adams3}
R.C.~Adams, J.M.~Cohen,
\href{https://adsabs.harvard.edu/pdf/1975ApJ...198..507A}{Astrophys. J. \textbf{198}, 507-512 (1975).}
\bibitem{penrose1}
R. Penrose, W. Rindler, ``Spinors and Space-Time (Cambridge Monographs on Mathematical Physics)" ,
\href{https://www.cambridge.org/core/books/spinors-and-spacetime/B66766D4755F13B98F95D0EB6DF26526}{ Cambridge: Cambridge University Press (1984).}
\bibitem{Coley:2017woz}
A.~A.~Coley, D.~D.~McNutt and A.~A.~Shoom,
\href{https://www.sciencedirect.com/science/article/pii/S0370269317303544?via}{Phys. Lett. B \textbf{771}, 131-135 (2017).}
\bibitem{Layden:2022oxu}
N.~T.~Layden, A.~A.~Coley and D.~D.~McNutt,
\href{https://link.springer.com/article/10.1007/s10714-022-02962-z}{Gen. Rel. Grav. \textbf{54}, no.8, 75 (2022).}

\bibitem{mcnuttpage_spi}
D.~D.~McNutt, D.~N.~Page,
\href{https://link.aps.org/doi/10.1103/PhysRevD.95.084044}{Phys Rev D 95, 084044 (2017)}

\bibitem{cartan-karlhede-alg}
Karlhede, A. 
\href{https://doi.org/10.1007/BF00771861}{Gen Relat Gravit 12, 693–707 (1980)}

\bibitem{kramer}
Kramer, D., Stephani, H., MacCallum, M., and Herlt, E. ``Exact solutions of Einstein's field equations",
\href{https://link.springer.com/article/10.1023/B:GERG.0000038634.90397.ea}{Cambridge University Press, Cambridge, UK, (1980).}
































\end{thebibliography}
\end{document}